\documentclass[twocolumn,tightenlines,showkeys,preprintnumbers,floatfix,amsmath,amssymb,superscriptaddress,prb]{revtex4-1}
\pdfoutput=1
\usepackage[T1]{fontenc}
\usepackage[utf8]{inputenc}
\usepackage{lmodern}
\usepackage{dcolumn}
\usepackage{bm}
\usepackage[english]{babel}
\usepackage[pdftex]{graphicx}
\usepackage{url}
\usepackage{breakurl}
\usepackage[breaklinks]{hyperref}

\newcommand*{\citen}{}
\DeclareRobustCommand*{\citen}[1]{%
	\begingroup
	\romannumeral-`\x 
	\setcitestyle{numbers}%
	\cite{#1}%
	\endgroup
}

\begin{document}

\title{Synthetic ferrimagnet spin transfer torque oscillator: model and non-linear properties}

\author{B. Lacoste}
	\affiliation{International Iberian Nanotechnology Laboratory, Braga, Portugal}
\author{M. Romera}\thanks{M. Romera is now working at Unité Mixte de Physique, CNRS, Thales, Univ. Paris-Sud, Université Paris-Saclay, 91767 Palaiseau, France}
	\affiliation{Univ. Grenoble Alpes, CEA, INAC-SPINTEC, CNRS, SPINTEC F-38000 Grenoble, France}
\author{U. Ebels}
	\affiliation{Univ. Grenoble Alpes, CEA, INAC-SPINTEC, CNRS, SPINTEC F-38000 Grenoble, France}
\author{L. D. Buda-Prejbeanu}
	\affiliation{Univ. Grenoble Alpes, CEA, INAC-SPINTEC, CNRS, SPINTEC F-38000 Grenoble, France}
\date{Date of submission \today}

\begin{abstract}
The non-linear parameters of spin-torque oscillators based on a synthetic ferrimagnet free layer (two coupled layers) are computed. The analytical expressions are compared to macrospin simulations in the case of a synthetic ferrimagnet excited by a current spin-polarized by an external fixed layer. It is shown that, of the two linear modes, acoustic and optical, only one is excited at a time, and therefore the self-sustained oscillations are similar to the dynamics of a single layer. However, the non-linear parameters values can be controlled by the parameters of the synthetic ferrimagnet. With a strong coupling  between the two layers and asymmetric layers (different thicknesses), it is demonstrated that the non-linear frequency shift can be reduced, which results in the reduction of the linewidth of the power spectral density. For a particular applied field, the non-linear parameter can even vanish; this corresponds to a transition between a red-shift and a blue-shift frequency dependence on the current and a linewidth reduction to the \emph{linear} linewidth value.
\end{abstract}

\maketitle

Spin transfer torque oscillators (STOs) have promising applications as high frequency microwave generators. A typical STO nano-pillar is composed of two magnetic layers separated by a metallic spacer or an isolating barrier. The magnetization of the first magnetic layer remains fixed in-plane or out-of-plane. It acts as a spin polarizer for the current flowing through the nano-pillar. The magnetization of the second layer can be driven into self-sustained oscillations by an applied DC current due to spin transfer torque (STT)~\cite{Kiselev2003,Rippard2004,Tsoi2004}. The oscillation of the free layer magnetization gives rise to a variation of the resistance of the pillar, so that an alternative voltage appears at its boundaries. For a single domain free layer, the generated microwave signal is typically in the GHz range. However, the large linewidth, in the order of tens of MHz, is an obstacle for functional devices.\\
In order to improve the STO characteristics, an accurate and simple model describing the dynamics is fundamental. For a single-layer (SL) STO, the general framework of non-linear auto-oscillators (NLAO), proved to be a particularly well adapted model~\cite{Slavin2008,Slavin2009,Kim2008,Tiberkevich2008}. Indeed, most of the features exhibited by experimental devices could be explained within this framework, such as the field and current dependence of the frequency, the broadened linewidth~\cite{Sierra2012}, but also synchronization to an external signal or to other STOs~\cite{Georges2008a}. More importantly, this model defines a key parameter for understanding the STO behavior: the non-linear amplitude-phase coupling parameter. By evaluating this non-linear parameter from the magnetic properties of the layer, it was found that the linewidth of the STO was reduced when applying a transverse field, for instance~\cite{Slavin2009}. However this model is confined to a SL free layer and some recent works studied STO devices where the free layer is composed of two coupled layers constituting a synthetic ferromagnet (SyF)~\cite{Houssameddine2010,Nagasawa2014}. Allegedly, the additional coupling energy would increase the magnetic stiffness and reduce the fluctuations. However, coupled systems are also more complicated to understand and a general analytical model is necessary to explain and define the important parameters of its dynamics. Typically, it would be useful to be able to calculate the non-linear amplitude-phase coupling parameter of a SyF-STO.\\
To answer this question, we propose to extend this framework, the NLAO model, to describe the dynamics of two coupled layers subjected to spin transfer torque. In order to treat the most general case, different coupling are included~: the Ruderman-Kittel-Kasuya-Yosida (RKKY) interaction, the dipolar coupling and the mutual STT. The non-adiabatic STT (or field-like torque) is also included, although its effect was found to be negligible in the particular configurations examined in this paper. It is fundamental in the dynamics of self-polarized STO~\cite{Seki2010,Jenkins2014}, though.\\
The NLAO theory is based on a change of coordinates to complex variables to represent the magnetization dynamics of the layers. The phase and amplitude of the complex variables describe the non-linear dynamics of the auto-oscillator. The validity of this approach is limited to quasi-conservative trajectories, for which the energy is almost constant, and to small oscillation amplitudes. Using common diagonalization techniques, the conservative part of the magnetization equation of motion is simplified to two terms~: a linear and a non-linear contribution. The dissipative part, which is supposed to be small compared to the conservative part, defines the equilibrium energy of the auto-oscillator by balancing the negative Gilbert damping and the positive STT.\\
The first and second parts of this work describe the steps to extract the auto-oscillator equation for two coupled SyF layers in the macrospin approximation. In the third and fourth part, we describe the dynamics of the SyF-STO defined by two coupled equations, so the STO can be described by two modes. However, only one of them is usually excited into steady-state at a time, so the SyF-STO is equivalent to a single-layer (SL) STO. This is an important result of this paper. The parameters of the single-mode SyF-STO are computed, especially the non-linear parameter, which is responsible for the frequency tunability, the large linewidth and the synchronization bandwidth. Another important result of this paper is the link between the vanishing of the non-linear parameter and the transition between a redshift and a blueshift regime. Finally in the fifth part, we study how to decrease the linewidth of a SyF-STO by changing the coupling strength and the thickness of the layers.

\section{Description of the system}
\subsection{Landau-Lifshitz-Gilbert-Slonczewski equation}
\begin{figure}[!t]
\centering
\includegraphics[width=\linewidth]{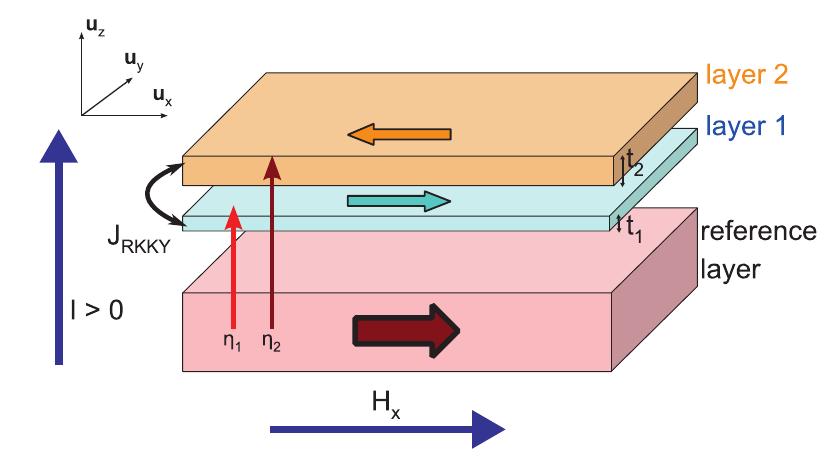}
\caption{Schematics of the synthetic ferrimagnet (SyF) as a free layer with a fixed in-plane magnetized reference layer.}
\label{fig:schema}
\end{figure}

We consider the system in Figure~\ref{fig:schema} of two magnetic layers, labeled 1 and 2 constituting a synthetic ferromagnet (SyF). The total free energy $E$ holds the demagnetizing energy, the uniaxial anisotropy energy and the Zeeman energy (including an exchange energy) of both layers, plus a conservative coupling term between the two layers, consisting of an RKKY interaction coupling and the dipolar coupling~:
\begin{align}
E&=\dfrac{\mu_0}{2} V_1 M_{1}H_{d1} (\bm{m}_1\cdot\bm{u_z})^2 -\dfrac{\mu_0}{2} V_1 M_{1}H_{k1} (\bm{m}_1\cdot\bm{u_x})^2 \nonumber\\
&\quad + \dfrac{\mu_0}{2} V_2 M_{2}H_{d2} (\bm{m}_2\cdot\bm{u_z})^2 -\dfrac{\mu_0}{2} V_2 M_{2}H_{k2} (\bm{m}_2\cdot\bm{u_x})^2 \nonumber\\
& \quad- \tilde{D}_x m_{1\,x} m_{2\,x} - \tilde{D}_y m_{1\,y} m_{2\,y} - \tilde{D}_z m_{1\,z} m_{2\,z} \nonumber\\
& \quad- \mu_0 V_1 M_{1}(H_{x}+ H_{\operatorname{ex}1})(\bm{m}_1\cdot\bm{u_x})  \nonumber\\
&\quad - \mu_0 V_2 M_{2}(H_{x}+ H_{\operatorname{ex}2})(\bm{m}_2\cdot\bm{u_x})
\label{eq:energy}
\end{align}
Here $\mu_0$ is the permeability of free space. $V_1=t_1 S$ and $V_2=t_2 S$ are the volumes of the layers, with thicknesses $t_1$ and $t_2$ and surface $S$. $M_1$ and $M_2$ are the saturation magnetizations of the layers, $H_{d1}$, $H_{d2}$ their demagnetizing fields (supposed positive), $H_{k1}$, $H_{k2}$ the uniaxial anisotropy fields. For each layer labeled by $i=(1,2)$, we define the demagnetizing coefficients $(N_{xx}^i, N_{yy}^i,N_{zz}^i)$, and the interface anisotropy constant $K_{S\,i}$, so that $H_{d\,i}=(N_{zz}^i-N_{yy}^i) M_i-2K_{S\,i}/(\mu_0 M_i t_i)$ and $H_{k\,i}=(N_{yy}^i-N_{xx}^i) M_i$. $H_{ex1}$, $H_{ex2}$ are the exchange fields acting on each layer (for instance from a coupling with a fixed anti-ferromagnet), and $H_x$ the applied field along the easy axis. \\
The coefficients $\tilde{D}_x$, $\tilde{D}_y$ and $\tilde{D}_z$ account for the (conservative) coupling between the two layers. They include RKKY interaction term and the dipolar coupling, such that, for $i=(x,y,z)$, $\tilde{D}_i = S J_{\text{RKKY}} + D_i$, where the $D_i$ are the dipolar coupling energy coefficients in macrospin.
Note that a negative $J_{\text{RKKY}}$ corresponds to an anti-ferromagnetic coupling between the layers. Two adjacent layers give rise to negative $D_x$ and $D_y$, and positive $D_z$.\\
\par

The layers are also subject to a spin transfer torque (STT) due to a current flowing perpendicular to the layers. A positive current corresponds to electrons flowing from layer 2 towards layer 1, and then to the reference layer. Thus, layer 1 is subjected to the STT from the reference layer and to the STT from layer 2 (with a negative factor because layer 1 receives reflected electrons from layer 2). Layer 2 is subjected to the STT from the reference layer and from layer 1 (because of electrons that were spin polarized after passing through layer 1). These spin torques acting on the two layers are modeled by two spin torque potentials, for layer 1 and 2 respectively, $P_1$ and $P_2$~\cite{Lacoste2014}:
\begin{align*}
P_1&=-\dfrac{\hbar}{2\lvert e\rvert}I\eta_1 \bm{m}_1\cdot\bm{u_x}+ \dfrac{\hbar}{2\lvert e\rvert}I\eta_{21}\bm{m}_1\cdot\bm{m}_2 \nonumber\\
P_2&=-\dfrac{\hbar}{2\lvert e\rvert}I\eta_2 \bm{m}_2\cdot\bm{u_x}- \dfrac{\hbar}{2\lvert e\rvert}I\eta_{12}\bm{m}_1\cdot\bm{m}_2
\end{align*}
$I$ is the current flowing through the layers. $\eta_1$ (resp. $\eta_2$) is the effective spin-polarization of the current in layer 1 (2) due to the fixed in-plane polarizer positioned before layer 1 according to the direction of the current. $\eta_{12}$ (resp. $\eta_{21}$) is the effective spin-polarization of the current in layer 2 (1) due to layer 1 (2).\\
Moreover, the two layers are subjected to perpendicular (or field-like) spin transfer torque (pSTT), from the reference layer and from the other layer. The pSTT is modeled by two potentials, similar to the spin torque potentials defined above~:
\begin{align*}
\tilde{P}_1&=-\dfrac{\hbar}{2\lvert e\rvert}I\beta_1 \bm{m}_1\cdot\bm{u_x}+ \dfrac{\hbar}{2\lvert e\rvert}I\beta_{21}\bm{m}_1\cdot\bm{m}_2 \nonumber\\
\tilde{P}_2&=-\dfrac{\hbar}{2\lvert e\rvert}I\beta_2 \bm{m}_2\cdot\bm{u_x}- \dfrac{\hbar}{2\lvert e\rvert}I\beta_{12}\bm{m}_1\cdot\bm{m}_2
\end{align*}

The equation of motion is given by the Landau-Lifshitz-Gilbert-Slonczewski (LLGS) equation. In this form, the damping is defined with respect to the time-derivative of the magnetization vector; after moving all the time-derivatives on the left-hand-side, the LLGS writes~:
\begin{align*}
\mu_0 V_1 M_{1}\dfrac{\text{d}\bm{m}_1}{\text{d}t}&=\gamma_0\bm{m}_1\times\dfrac{\partial E}{\partial \bm{m}_1} + \gamma_0\bm{m}_1\times  \dfrac{\partial \tilde{P}_1}{\partial \bm{m}_1} \\
&\qquad + \gamma_0\bm{m}_1\times\bigg(\bm{m}_1\times\dfrac{\partial}{\partial \bm{m}_1}(P_1-\alpha_1 E)\bigg)\nonumber\\
\mu_0 V_2 M_{2}\dfrac{\text{d}\bm{m}_2}{\text{d}t}&=\gamma_0\bm{m}_2\times  \dfrac{\partial E}{\partial \bm{m}_2} + \gamma_0\bm{m}_2\times  \dfrac{\partial \tilde{P}_2}{\partial \bm{m}_2} \\
&\qquad + \gamma_0\bm{m}_2\times\bigg(\bm{m}_2\times  \dfrac{\partial }{\partial \bm{m}_2} (P_2-\alpha_2 E)\bigg)  \nonumber
\end{align*}
The Gilbert damping coefficients of the two layers are given by $\alpha_1$ and $\alpha_2$. The correction to the gyromagnetic ratio due to the damping coefficient has been neglected, so $\gamma_0=\mu_0 \gamma$ where $\gamma$ is the gyro-magnetic ratio.\\
According to the form of the LLGS equation used in this paper, the coefficients $\beta_j$ ($j=(1,2,12,21)$) of the field-like torques can contain a term proportional to the coefficients $\eta_j$ from the damping-like STT and to the Gilbert damping constants of the two layers, that we call pseudo-field-like torque. Namely $\beta_1 = \alpha_1 \eta_1$,  $\beta_2 = \alpha_2 \eta_2$,  $\beta_{12} = \alpha_2 \eta_{12}$ and $\beta_{21} = \alpha_1 \eta_{21}$. Such additional terms would be coming from the transformation of the STT from the Gilbert-form of the LLGS equation to the Landau-form.\\

By writing the LLGS equation in this form, the free energy part, which is common to both layers, is separated from the rest. This will allow to use a similar formalism as for the description of a single layer in previous publications~\cite{Slavin2008,Slavin2009}.\\

In order to simplify the notations, we introduce the layer asymmetry $\beta$, the geometrical mean magnetic volume $\mathcal{M}$ and the following normalized hamiltonian and potentials~:
\begin{align*}
\beta &= \sqrt{\dfrac{M_2 t_2}{M_1 t_1}} & \mathcal{M} &= \mu_0 S \sqrt{M_1 t_1 M_2 t_2}
\end{align*}
\begin{align*}
\mathcal{H} &= \dfrac{\gamma_0 E}{2\mathcal{M}} & \Delta_1 &= \dfrac{\gamma_0 \tilde{P}_1}{2\mathcal{M}} & \Delta_2 &= \dfrac{\gamma_0 \tilde{P}_2}{2\mathcal{M}}
\end{align*}
\begin{align*}
\Gamma_1 &= \alpha_1 \mathcal{H} - \dfrac{\gamma_0 P_1}{2\mathcal{M}} & \Gamma_2 &= \alpha_2 \mathcal{H} - \dfrac{\gamma_0 P_2}{2\mathcal{M}}
\end{align*}

In the following, dotted variables represent their time derivative. Therefore the LLGS equation rewrites~:
\begin{align}\label{eq:LL}
\dfrac{1}{2\beta} \dot{\bm{m}_1} &= \bm{m}_1 \times \dfrac{\partial \mathcal{H}}{\partial \bm{m}_1} + \bm{m}_1 \times \dfrac{\partial \Delta_1}{\partial \bm{m}_1} \nonumber\\
&\quad + \bm{m}_1 \times \left(\bm{m}_1 \times \dfrac{\partial \Gamma_1}{\partial \bm{m}_1} \right) \\
\dfrac{\beta}{2} \dot{\bm{m}_2} &= \bm{m}_2 \times \dfrac{\partial \mathcal{H}}{\partial \bm{m}_2} + \bm{m}_2 \times \dfrac{\partial \Delta_2}{\partial \bm{m}_2} \nonumber\\
&\quad+ \bm{m}_2 \times \left(\bm{m}_2 \times \dfrac{\partial \Gamma_2}{\partial \bm{m}_2} \right) \nonumber 
\end{align}
We decompose the right-hand-side of the LLGS equation in two parts that will be treated separately: (i) the conservative hamiltonian terms (simply called conservative in the following) that are composed of the first terms on the right-hand-side and depend only on $\mathcal{H}$. (ii) the conservative non-hamiltonian terms and the dissipative terms (simply called dissipative in the following because the dissipative terms play a more important role) that are composed of the other two terms (respectively) on the right-hand-side.\\
In general, the damping constants $\alpha_1, \alpha_2$ are considered to be small ($<0.1$) and the applied current is reasonably small, so the conservative part is larger than the dissipative part. The two different orders of magnitude further support the distinction made between the two parts.
\par
\subsection{Numerical parameters}\label{section:parameters}

The results from the extended NLAO model will be compared to macrospin LLGS simulations. The case of an asymmetric SyF shows interesting properties, especially in terms of linewidth reduction. As all cases cannot be reproduced here, we focus on a SyF with thickness asymmetry between the two layers. However, asymmetry can also be introduced by submitting one layer to an exchange field or by reducing the effective demagnetizing field of one of the layers with perpendicular interface anisotropy.\\
For the inter-layer coupling, two regimes are considered, small coupling $J_{\operatorname{RKKY}}=-2\times10^{-4}$~J/m$^2$ and large coupling $J_{\operatorname{RKKY}}=-5\times10^{-4}$~J/m$^2$. The dipolar coupling is neglected in the macrospin simulations. This is supported by the fact that, in the macrospin approximation and in nano-pillars with circular cross-section, the dipolar coupling is an antiferromagnetic coupling in the in-plane directions ($x$ and $y$~directions in our convention) and a ferromagnetic coupling in the normal direction ($z$~direction). Because of the high demagnetizing field in thin layers, the trajectories have a small out-of-plane component, so contribution from the dipolar coupling is comparable to a low RKKY antiferromagnetic coupling. For the layer thicknesses considered, the dipolar field is lower than the RKKY coupling field, so the dipolar coupling is simply neglected.

The rest of the parameters are defined in Table~\ref{table:macrospin}.\\
According to the value of the area $S$ of the pillars, currents expressed in mA correspond to current densities of $10^{11}$~A/m$^2$.\\
The current is considered to be unpolarized after going through the first layer, so $\eta_2=0$. However the same qualitative results were obtained~\cite{Ichimura2011} if we suppose that $\eta_2=\pm\eta_1$.
\begin{table}[!htb]
 	\renewcommand{\arraystretch}{1.7}
	\begin{center}
	\resizebox{0.7\linewidth}{!} {
		\begin{tabular}{ c  c  c }
		
		 \hline\hline
		 Identical properties & Value  \\ \hline
		 $M_{s1},M_{s2}$			& $1\times 10^{6}$~A/m  \\ 
		 $H_{d1},H_{d2}$			& $0.9\times 10^{6}$~A/m  \\ 
		 $H_{k1},H_{k2}$			& $10\times 10^{3}$~A/m  \\ 
		 $H_{\text{ex}\,1},H_{\text{ex}\,2}$  & 0  \\ 
		 $\alpha_1,\alpha_2$		& 0.02 \\
		 $S$ 						& $10^{-14}$~m$^2$ \\
		 $\eta_{21}, \eta_{12}$     & 0	\\
		 $\beta_1,\beta_2,\beta_{12},\beta_{21}$ & 0 \\
		 \hline	Different properties & Values  \\ \hline
		 $t_1$ and $t_2$ 					& 1.8 and 2.2~nm\\ 
		 $\eta_1$ and $\eta_2$  			& 0.5 and 0 \\
		 \hline\hline
		\end{tabular} 
	}
	\renewcommand{\arraystretch}{1.5}\\[1em]
	\end{center}
	\caption{Properties of the magnetic layers.}
	\label{table:macrospin}
\end{table}
\par
The conservative part is the most important to describe the self-sustained oscillations because the trajectories of the self-sustained oscillations are close to the constant energy trajectories. For this reason, a change of variables that describes accurately the conservative part and treats the dissipative part as a small perturbation is adapted to describe the dynamics of the STO. This will be developed in the next part.

\section{Transformation to complex variables}
\subsection{Complex variables: conservative part}\label{section:transform}
We intend to rewrite the LLGS equation in complex form representing the evolution of two modes $a_1$ and $a_2$. Let $\mathbf{a}=(a_1,a_2)$ be a 2-dimensional complex vector. The goal is to write the conservative part of the LLGS equation in the form~:
\begin{equation}
\dot{\mathbf{a}}=-i\dfrac{\partial \mathcal{H}}{\partial \mathbf{a}^{\dagger}}
\label{eq:form-a}
\end{equation}

The elements of the basis, $a_1$ and $a_2$, represent uniform modes around the equilibrium position, with complex conjugates $\mathbf{a}^{\dagger}=(a_1^{\dagger},a_2^{\dagger})$. In the following, we focus only on the modes around the parallel equilibrium state (or antiparallel depending on the sign of the RKKY coupling constant and the dipolar coupling), i.e. the synthetic ferrimagnet (SyF) is in the plateau region. The equilibrium position is represented by~:
\begin{align*}
m_{1\,x}^{\text{eq}} &= m \bm{u}_x  &  m_{2\,x}^{\text{eq}} &= m n \bm{u}_x
\end{align*}

Here $n=\operatorname{sign}(\tilde{D}_x)$ reflects the ferromagnetic or anti-ferromagnetic type of coupling between the two layers: $n=+1$ ferromagnetic coupling, $n=-1$ anti-ferromagnetic coupling. The direction of layer 1 relatively to the fixed reference layer is given by $m$~: $m=+1$ for a parallel (P) orientation, $m=-1$ for an antiparallel (AP) orientation. The initial state is then defined by a P or AP configuration (with respect to the reference layer) and a ferromagnetic or anti-ferromagnetic coupling between layer 1 and layer 2.\\
We proceed to a change of coordinate system so that the equilibrium magnetizations have the same definitions for all the layers. They are defined by $\bm{m}_{i}^{\text{eq}} = \bm{u}_{\zeta}^i$ for $i=(1,2)$~:
\begin{align*}
\bm{u}_{\zeta}^1 &= m \,\bm{u}_x &   \bm{u}_{\zeta}^2 &= m n \,\bm{u}_x \\
\bm{u}_{\xi}^1 &= m \,\bm{u}_y   &   \bm{u}_{\xi}^2 &= m n \,\bm{u}_y\\
\bm{u}_{\eta}^1 &= \bm{u}_z   & \bm{u}_{\eta}^2 &= \bm{u}_z
\end{align*}

The expressions of $a_1$ and $a_2$ with respect to the local magnetization coordinates have to be chosen adequately so that the conservative part of LLGS in this new system of coordinates take the hamiltonian form of Eq.~\eqref{eq:form-a}. For that we set~:
\begin{align}
a_1 = \dfrac{1}{\sqrt{\beta}}\dfrac{m_{1\,\xi} - i m_{1\,\eta}}{\sqrt{2(1+m_{1\,\zeta})}} \\
a_2 = \sqrt{\beta}\dfrac{m_{2\,\xi} - i m_{2\,\eta}}{\sqrt{2(1+m_{2\,\zeta})}} 
\end{align}
Notice that there are other choices of $(\mathcal{H}, a_1, a_2)$ that allows to rewrite the LLGS equation in the hamiltonian form of Eq.~\eqref{eq:form-a}, notably by multiplying $a_1$ and $a_2$ by the same constant term $C$ and $\mathcal{H}$ by $C^2$. The quadratic part (as it will be defined later) of the Hamiltonian would remain unchanged by changing this factor, but the quartic (and the other orders) part would be affected. Hence, it is not possible to compare coefficients of quartic or higher order for different geometries, as their definition depends on the arbitrary choice of the constant $C$. Instead, normalized coefficients should be compared.\\
The expression of $\mathcal{H}$ with respect to the new variables $(a_1, a_2)$ and their complex conjugates $(a_1^\dagger, a_2^\dagger)$ can be divided as $\mathcal{H} = \mathcal{H}_2 + \mathcal{H}_4$ by dropping the constant term and neglecting higher order hamiltonian terms. In terms of the complex variables, $\mathcal{H}_2$ is the quadratic part and $\mathcal{H}_4$ is the quartic part. 
\begin{align*}
\mathcal{H}_2 &= \mathcal{A}_1 a_1 a_1^{\dagger} + \mathcal{A}_2 a_2 a_2^{\dagger} + \dfrac{1}{2} \big(\mathcal{B}_1 a_1^2 + \mathcal{B}_2 a_2^2  + \mbox{c.c.} \big) \\
&\quad +  \big( \mathcal{C}_{12} a_1 a_2 + \mathcal{D}_{12} a_1 a_2^{\dagger} + \mbox{c.c.}  \big) \\
\mathcal{H}_4 &= \mathcal{U}_1 a_1^2 a_1^{\dagger 2} + \mathcal{U}_2 a_2^2 a_2^{\dagger 2} +  \mathcal{W}_{12}a_1 a_2 a_1^{\dagger} a_2^{\dagger} \\
&\quad  + \big(\mathcal{V}_1 a_1^3 a_1^{\dagger} + \mathcal{V}_2 a_2^3 a_2^{\dagger}  + \mbox{c.c.} \big) \\
&\quad   + \big( \mathcal{Y}_{12} a_1^2 a_1^{\dagger} a_2 + \mathcal{Y}_{21} a_1 a_2^2 a_2^{\dagger} + \mbox{c.c.}  \big) \\
&\quad  + \big( \mathcal{Z}_{12} a_1 a_1^{\dagger 2} a_2 + \mathcal{Z}_{21} a_1^{\dagger} a_2^2 a_2^{\dagger} + \mbox{c.c.}  \big) \\
\end{align*}

We introduce new parameters that correspond to the characteristic frequencies~:
\begin{align*}
\omega_k^1&=\gamma_0 H_{k\,1}\text{,}\quad \omega_k^2=\gamma_0 H_{k\,2}\text{,}\quad \omega_d^1 =\gamma_0 H_{d\,1}\text{,}\quad \omega_d^2 =\gamma_0 H_{d\,2}\text{,} \\
\omega_a^1&=\gamma_0 m(H_x + H_{\text{ex}\,1})\text{,}\quad \omega_a^2=\gamma_0 n m(H_x + H_{\text{ex}\,2}) \text{,} \\
\omega_c^0&=\dfrac{\gamma_0}{\mathcal{M}} n \tilde{D}_x\text{,}\quad
\omega_c^{-}=\dfrac{\gamma_0}{\mathcal{M}}  \dfrac{n\tilde{D}_y - \tilde{D}_z}{2}\text{,}\quad
\omega_c^{+}=\dfrac{\gamma_0}{\mathcal{M}}  \dfrac{n\tilde{D}_y + \tilde{D}_z}{2}\text{.}
\end{align*}
Using these notations, the coefficients of the hamiltonian are given by~: 
\begin{align*}
\mathcal{A}_1 &= \omega_k^1+\dfrac{\omega_d^1}{2}+ \omega_a^1+\beta\omega_c^0 \text{,}\quad
\mathcal{A}_2 = \omega_k^2+\dfrac{\omega_d^2}{2}+ \omega_a^2+\dfrac{\omega_c^0}{\beta}\text{,} \\
\mathcal{B}_1 &= -\dfrac{\omega_d^1}{2}\text{,}\quad  \mathcal{B}_2 = -\dfrac{\omega_d^2}{2} \text{,}\quad \mathcal{C}_{12} = - \omega_c^{-} \text{,}\quad \mathcal{D}_{12} = - \omega_c^{+} \text{,} \\
\mathcal{U}_1 &= -\beta\omega_k^1-\dfrac{\beta}{2}\omega_d^1 \text{,}\quad
\mathcal{U}_2 = -\dfrac{\omega_k^2}{\beta}-\dfrac{\omega_d^2}{2 \beta} \text{,}\\ \mathcal{W}_{12} &= -2 \omega_c^0 \text{,}\quad
\mathcal{V}_1 = \dfrac{\beta}{4}\omega_d^1 \text{,}\quad
\mathcal{V}_2 = \dfrac{\omega_d^2}{4\beta} \text{,} \\
\mathcal{Y}_{12} &= \dfrac{\beta}{2}\omega_c^{-} \text{,}\quad
\mathcal{Y}_{21} = \dfrac{\omega_c^{-}}{2\beta} \text{,}\quad
\mathcal{Z}_{12} = \dfrac{\beta}{2}\omega_c^{+}  \text{,}\quad
\mathcal{Z}_{21} = \dfrac{\omega_c^{+}}{2\beta} \text{.}\\
\end{align*}

In matrix form, $\mathcal{H}_2$ rewrites~:
\begin{align*}
\mathcal{H}_2 &= \dfrac{1}{2}\begin{pmatrix} a_1^{\dagger} & a_2^{\dagger} & a_1 & a_2 \end{pmatrix}\begin{pmatrix}
\mathcal{A}_1 & \overline{\mathcal{D}}_{12} & \overline{\mathcal{B}}_{1} & \overline{\mathcal{C}}_{12} \\
\mathcal{D}_{12} & \mathcal{A}_2 & \overline{\mathcal{C}}_{12} & \overline{\mathcal{B}}_{2} \\
\mathcal{B}_1 & \mathcal{C}_{12} & \mathcal{A}_{1} & \mathcal{D}_{12} \\
\mathcal{C}_{12} & \mathcal{B}_2 & \overline{\mathcal{D}}_{12} & \mathcal{A}_{2}
\end{pmatrix}\begin{pmatrix} a_1 \\ a_2 \\  a_1^{\dagger} \\ a_2^{\dagger} \end{pmatrix}
\end{align*}

The notation $\overline{x}$ is used for the complex conjugate of $x$, to distinguish scalar coefficients from the magnetization complex variables $(a_1, a_2)$ with complex conjugates $(a_1^\dagger, a_2^\dagger)$. As for a SL oscillator, it is possible to diagonalize the quadratic part $\mathcal{H}_2$ of the hamiltonian~\cite{Hemmen1980}. In fact it is possible to do so for any number of layers, although it becomes difficult to find analytical expressions for more than two layers. The new complex basis is called $(b_{\operatorname{op}},b_{\operatorname{ac}})$, with~:
\begin{align}\label{eq:T-transform}
\begin{pmatrix} a_1 \\ a_2 \\  a_1^{\dagger} \\ a_2^{\dagger} \end{pmatrix} &= T_{ab} \begin{pmatrix} b_{\operatorname{op}} \\ b_{\operatorname{ac}} \\  b_{\operatorname{op}}^{\dagger} \\ b_{\operatorname{ac}}^{\dagger} \end{pmatrix}
\end{align}

\begin{align*}
T_{ab} &= \begin{pmatrix}
u_1^{\operatorname{op}} & u_1^{\operatorname{ac}} & v_1^{\operatorname{op}} & v_1^{\operatorname{ac}} \\
u_2^{\operatorname{op}} & u_2^{\operatorname{ac}} & v_2^{\operatorname{op}} & v_2^{\operatorname{ac}} \\
\overline{v}_1^{\operatorname{op}} & \overline{v}_1^{\operatorname{ac}} & \overline{u}_1^{\operatorname{op}} & \overline{u}_1^{\operatorname{ac}} \\
\overline{v}_2^{\operatorname{op}} & \overline{v}_2^{\operatorname{ac}} & \overline{u}_2^{\operatorname{op}} & \overline{u}_2^{\operatorname{ac}} \\
\end{pmatrix}
\end{align*}
We note $\widehat{I}$ the 4x4 block matrix $\widehat{I}=\begin{pmatrix}I_2 & 0 \\ 0 & -I_2 \end{pmatrix}$ with $I_2$ the $2\times2$ unity matrix. $T_{ab}^{\dagger}$ is the transpose conjugate of $T_{ab}$. It verifies~:
\begin{align*}
T_{ab}^{-1} &= \widehat{I} T_{ab}^{\dagger}\widehat{I} 
\end{align*}
In the new basis, $\mathcal{H}_2$ takes the simple form~:
\begin{align*}
\mathcal{H}_2 &= \omega_{\operatorname{op}} b_{\operatorname{op}} b_{\operatorname{op}}^{\dagger}  + \omega_{\operatorname{ac}} b_{\operatorname{ac}} b_{\operatorname{ac}}^{\dagger}
\end{align*}
The complex variables $(b_{\operatorname{op}},b_{\operatorname{ac}})$ are eigenvectors of the linear hamiltonian. They correspond to the two linear modes of the SyF-STO: optical and acoustic. This base of eigenvectors is then used to express the non-linear part of the hamiltonian.\par
The expressions of $\omega_{\text{op}}$, $\omega_{\text{ac}}$ and of the coefficients of the matrix $T_{ab}$ come from diagonalizing the matrix $\tilde{\mathcal{H}}_2$~:
\begin{align*}
\tilde{\mathcal{H}}_2 &= \begin{pmatrix}
\mathcal{A}_1 & \overline{\mathcal{D}}_{12} & \overline{\mathcal{B}}_{1} & \overline{\mathcal{C}}_{12} \\
\mathcal{D}_{12} & \mathcal{A}_2 & \overline{\mathcal{C}}_{12} & \overline{\mathcal{B}}_{2} \\
-\mathcal{B}_1 & -\mathcal{C}_{12} & -\mathcal{A}_{1} & -\mathcal{D}_{12} \\
-\mathcal{C}_{12} & -\mathcal{B}_2 & -\overline{\mathcal{D}}_{12} & -\mathcal{A}_{2}
\end{pmatrix}
\end{align*}
The expression of $\tilde{\mathcal{H}}_2$ is for the general case, for any direction of the equilibrium magnetizations. In the configuration studied here, with equilibrium configurations and applied fields along the easy axis, all the coefficients are real. We take this assumption in the following.\\
From computing the eigenvalues of $\tilde{\mathcal{H}}_2$, the following values are obtained for $\omega_{\text{op}/\text{ac}}$~:
\begin{align*}
\omega_{\text{av}}^2 &= (\mathcal{A}_1^2+\mathcal{A}_2^2)/2. - (\mathcal{B}_1^2+\mathcal{B}_2^2)/2. + \mathcal{D}_{12}^2 - \mathcal{C}_{12}^2 \\
\Delta &= \Big(\mathcal{A}_1^2 - \mathcal{A}_2^2 - \mathcal{B}_1^2 + \mathcal{B}_2^2\Big)^2 \\
&\quad + 4\Big(\mathcal{C}_{12}(\mathcal{B}_1+\mathcal{B}_2) - \mathcal{D}_{12}(\mathcal{A}_1+\mathcal{A}_2)\Big)^2 \\
&\quad - 4\Big(\mathcal{C}_{12}(\mathcal{A}_1-\mathcal{A}_2) - \mathcal{D}_{12}(\mathcal{B}_1-\mathcal{B}_2)\Big)^2 
\end{align*}
\begin{align}
\omega_{\operatorname{op/ac}}^2 &= \omega_{\text{av}}^2 \pm \dfrac{\sqrt{\Delta}}{2}
\end{align}
The frequencies $\omega_{\text{op}}$ and $\omega_{\text{ac}}$ correspond to the two modes optical and acoustic that are observed in ferromagnetic resonance (FMR) experiments with SyFs. By definition, the optical mode corresponds to the mode with the highest frequency. The expressions of the two mode frequencies are in agreement with the expressions found in the literature~\cite{Devolder2012, Lacoste2014}.

The eigenvectors of $\tilde{\mathcal{H}}_2$, which correspond to the columns of the matrix $T_{ab}$, have complicated expressions. However, due to normalization conditions, they can be expressed by 6 angles. For the two labels $j=(\text{op},\text{ac})$, the elements of the matrix $T_{ab}$ are given by~:
\begin{align*}
u_1^{j} &=  \cosh \theta_j \cos \phi_j \\
u_2^{j} &= - \cosh \theta_j \sin \phi_j \\
v_1^{j} &= - \sinh \theta_j \cos \psi_j \\
v_2^{j} &=  \sinh \theta_j \sin \psi_j 
\end{align*}

The details about the coefficients are given in Appendix~\ref{A:transformationT}.\\
The angles $\phi_j$ and $\psi_j$ are related to the coupling between the two layers. In fact, if the coupling vanishes ($\mathcal{C}_{12}=\mathcal{D}_{12}=0$), these angles vanish for one mode, say the acoustic mode, $\phi_{\text{ac}} = \psi_{\text{ac}} = 0 $, whereas for the other mode, $\phi_{\text{op}} = \psi_{\text{op}} = \pi/2$. So the optical mode $b_{\text{op}}$ depends only on the layer 2 complex variable $a_2$ and the acoustic mode $b_{\text{ac}}$ on the layer 1 and $a_1$.\\
The angles $\theta_j$ correspond to the mixing between the diagonal terms $\mathcal{A}_1$, $\mathcal{A}_2$ and the off-diagonal terms $\mathcal{B}_1$, $\mathcal{B}_2$, by analogy to the transformation coefficients for a single layer.\\
\par

However, it is not possible to obtain an exact diagonalization of the quartic part $\mathcal{H}_4$ of the hamiltonian but non-canonical transformations provide good approximations. We distinguish the resonant terms, for which the overall phase vanishes, like $b_{\text{op}}b_{\text{op}}^{\dagger}$, from the non-resonant (or off-diagonal) terms, for which the overall phase varies with time, like $b_{\text{op}}b_{\text{ac}}^{\dagger}$.\\
Because in this configuration, all along the easy axis, there is no cubic term in the Hamiltonian, the (non-canonical) transformation to remove the conservative non-resonant terms~\cite{Lvov1994} does not affect the value of the diagonal quartic terms. Equivalently, we then assume that the off-diagonal terms of the quartic term are negligible. However this assumption is valid only if the mode frequencies $\omega_{\text{op}}$ and $\omega_{\text{ac}}$ are large compared to the off-diagonal terms. Concretely, when applying an external field that is comparable to the spin-flop field, the acoustic mode frequency almost vanishes and the previous assumption is no longer valid. In this case the dynamics is more complicated because the conservative non-resonant terms become important. We therefore limit the major discussion to the field range below the spin-flop field.\\
Neglecting the non-resonant terms, the quartic part has the simple expression~:
\begin{align*}
\mathcal{H}_4 &= \dfrac{N_{\text{op}}}{2} b_{\text{op}}^2 {b_{\text{op}}^{\dagger}}^2  + \dfrac{N_{\text{ac}}}{2} b_{\text{ac}}^2 {b_{\text{ac}}^{\dagger}}^2  + T b_{\text{op}}b_{\text{op}}^{\dagger}b_{\text{ac}} b_{\text{ac}}^{\dagger}
\end{align*}
Where $N_{\textrm{ac}}$ (resp. $N_{\textrm{op}}$) is the acoustic (optical) non-linear frequency shift coefficient and $T$ is the mixed-mode non-linear frequency shift coefficient. They are all real.\\
All these coefficients come from the conservative part of the LLGS equation, they depend on the demagnetizing fields of the layers, applied field and coupling energy. However, they are independent of the damping coefficients of the layers and of the applied current.\\

\subsection{Complex variables: dissipative part}
We now focus on the dissipative part of the LLGS equation. After the transformation to the complex variables $a_1,a_2$, the LLGS equation writes~:
\begin{equation}
\dot{\mathbf{a}}=-i\dfrac{\partial \mathcal{H}}{\partial \mathbf{a}^{\dagger}} - \mathbf{F}_a
\label{eq:form-a-complete}
\end{equation}
Where $\mathbf{F}_a = (F_{a_1}, F_{a_2})$ is a vector with two complex components. The two dissipative complex components $F_{a_1}, F_{a_2}$ are truncated to contain only linear and cubic terms in $a_1$, $a_2$, $a_1^{\dagger}$ and $ a_2^{\dagger}$. The polynomial coefficients are noted with 4 indices $(k,l,m,n)$, so that~:
\begin{align*}
F_{a_i} = \sum_{k,l,m,n} f_{a_i}^{k,l,m,n} {a_1}^k{a_2}^l{a_1^{\dagger}}^m{a_2^{\dagger}}^n \qquad \mbox{for }i=1,2
\end{align*}
The expressions of the coefficients of the dissipative terms are given in the Appendix~\ref{A:dissip}.
Using the linear transform with the matrix $T_{ab}$, similar coefficients for the $b$-variables are obtained, with $\mathbf{b} = (b_{\text{op}}, b_{\text{ac}})$~:
\begin{align*}
\dot{\mathbf{b}} &=-i\dfrac{\partial \mathcal{H}}{\partial \mathbf{b}^{\dagger}} - \mathbf{F}_b \\
\mathbf{F}_b &= (F_{b_{\text{op}}}, F_{b_{\text{ac}}}) \qquad \mbox{and for $i=$~op,~ac~:}\\
F_{b_i} &= \sum_{k,l,m,n} f_{b_i}^{k,l,m,n} {b_{\text{op}}}^k{{b_{\text{ac}}}}^l{{b_{\text{op}}}^{\dagger}}^m{{b_{\text{ac}}}^{\dagger}}^n
\end{align*}
So that~:
\begin{align*}
\begin{pmatrix}\mathbf{F}_b \\ \mathbf{F}_b^{\dagger}\end{pmatrix} 
&= T_{ab}^{-1} \cdot \begin{pmatrix}\mathbf{F}_a \\ \mathbf{F}_a^{\dagger}\end{pmatrix}
\end{align*}
Where $\mathbf{F}_a$ and $\mathbf{F}_a^{\dagger}$ are expressed in terms of $b$-variables using the transform of equation~\eqref{eq:T-transform}.\\

All these $f_{b_i}^{k,l,m,n}$ coefficients in the $b$-coordinates are complex in general. However, if the coefficients of the conservative terms are real ($\mathcal{B}_1$, $\mathcal{B}_2$, $\mathcal{C}_{12}$, $\mathcal{D}_{12}$, $\mathcal{V}_1$, $\mathcal{V}_2$, etc...), only the field-like torque contributes to the imaginary part. For the magnetic configuration studied in this paper, the applied field is aligned with the magnetization, so the conservative coefficients are real. The field-like torque is also set to zero, so the dissipative coefficients are real. In any case, the real part is the really important part, as it defines the power as it will be shown in the next section. The imaginary part only gives a contribution to the phase equation and it is negligible compared to the contribution from the conservative part in the configuration studied in this paper, with an external polarizer. Without external polarizer, but taking into account the mutual spin-torque in a self-polarizer structure, the contribution of the field-like torque is non-negligible as shown in reference~\citen{Romera2016}.\\
\par
Of all the dissipative terms, the most important are the resonant terms, i.e. the terms that are similar to the resonant terms from the conservative part. Taking into account only these resonant terms, the dissipative part reduces to~:
\begin{align}\label{eq:dissip_final}
F_{b_{\text{op}}} &= b_{\text{op}} \Big( \gamma_{\text{op}} + Q_{\text{op}}b_{\text{op}}b_{\text{op}}^{\dagger} + R_{\text{op}}b_{\text{ac}}b_{\text{ac}}^{\dagger}\Big) \nonumber\\
F_{b_{\text{ac}}} &= b_{\text{ac}} \Big( \gamma_{\text{ac}} + Q_{\text{ac}}b_{\text{ac}}b_{\text{ac}}^{\dagger} + R_{\text{ac}}b_{\text{op}}b_{\text{op}}^{\dagger}\Big) 
\end{align}
$\gamma_{\textrm{ac}}$ (resp. $\gamma_{\textrm{op}}$) is the acoustic (optical) linear relaxation rate. $Q_{\textrm{ac}}$ ($Q_{\textrm{op}}$) is the acoustic (optical) non-linear relaxation rate coefficient. $R_{\textrm{ac}}$ ($R_{\textrm{op}}$) is the coefficient of the acoustic (optical) non-linear mode mixing relaxation rate.\\
Because of the linear dependence of the STT amplitude with respect to the applied current hypothesized in this paper, these coefficients depend linearly on the applied current. The linear coefficients, $\gamma_{\textrm{op}}$ and $\gamma_{\textrm{ac}}$, are positive for zero current, in agreement with the fact that the Gilbert damping is a relaxation to the minimum energy configuration. They decrease with the current if the current is applied in the direction that destabilizes the magnetization. The dissipative coefficients also depend on the demagnetizing fields and coupling energy, like the conservative coefficients.\\
\par
The analytical expressions of the coefficients are very lengthy and therefore they are not presented here in detail. Instead, for each value of field and current, the coefficients are calculated numerically through the various transformations, using the materials parameters given in section~\ref{section:parameters}. The variation of the different coefficients with field and current are given in section~\ref{section:simple_dynamics}, where the coupled complex equations are solved.

\section{Dynamics with resonant terms only}\label{section:simple_dynamics}
In order to illustrate some of the basic features of the coupled system, in a first approximation only the resonant terms, i.e. $\mathcal{H}_2$, $\mathcal{H}_4$ and Eq.~\ref{eq:dissip_final} are considered for the time evolution of $b_{\textrm{op}}$ and $b_{\textrm{ac}}$:
\begin{align}\label{eq:coupled_1}
\dot{b}_{\operatorname{op}} &= -ib_{\operatorname{op}}(\omega_{\operatorname{op}}+ N_{\operatorname{op}}p_{\operatorname{op}} + T p_{\operatorname{ac}}) \nonumber\\
  & \quad - b_{\operatorname{op}}( \gamma_{\operatorname{op}} + Q_{\operatorname{op}} p_{\operatorname{op}} + R_{\operatorname{op}} p_{\operatorname{ac}}) \nonumber\\
\dot{b}_{\operatorname{ac}} &= -i b_{\operatorname{ac}}(\omega_{\operatorname{ac}}+ N_{\operatorname{ac}}p_{\operatorname{ac}} + T p_{\operatorname{op}}) \nonumber\\
  & \quad - b_{\operatorname{ac}}( \gamma_{\operatorname{ac}} + Q_{\operatorname{ac}} p_{\operatorname{ac}} + R_{\operatorname{ac}} p_{\operatorname{op}})
\end{align}
Where $p_{\operatorname{op}}=b_{\text{op}}b_{\text{op}}^{\dagger}$ and $p_{\operatorname{ac}}=b_{\text{ac}}b_{\text{ac}}^{\dagger}$ are the powers of the two modes. All the coefficients are supposed to be real.\\
We notice that the dynamics of the coupled system does not reduce to two independent oscillator equations. Even if the two modes are decoupled in the linear regime ($p_{\operatorname{op}}, p_{\operatorname{ac}}\ll 1$), the acoustic and optical modes are coupled through the non-linear coefficients.\\
Introducing the phases $\phi_{\operatorname{op}}, \phi_{\operatorname{ac}}$ of the two modes, let's define~:
\begin{align*}
b_{\operatorname{op}} &= \sqrt{p_{\operatorname{op}}}e^{-i\phi_{\operatorname{op}}}\\
b_{\operatorname{ac}} &= \sqrt{p_{\operatorname{ac}}}e^{-i\phi_{\operatorname{ac}}}
\end{align*}

Using the definitions of $b_{\textrm{op}}$ and $b_\textrm{ac}$, one can derive separate equations for the power and the phase. These will be discussed in the next sections. It is reminded that for a single layer the equivalent analytical equations yield as a stationary solution a constant oscillation power (cancellation of the dissipative part). In the next section it is shown that the coupled Eq.~\ref{eq:coupled_1} can reduce to a single mode equation under specific conditions. For this we start discussing the solutions to the power equations.\\
\par

\subsection{Power equations}\label{section:power-equation}
The equations of time evolution of the power and phase are derived from the complex equations~\ref{eq:coupled_1}. The equations of evolution of the powers of both modes are given by the generalized Lotka-Volterra (LV) equations~\cite{DeAguiar2007}:
\begin{align}\label{eq:powers}
\dot{p}_{\operatorname{op}} &= -2 p_{\operatorname{op}}( \gamma_{\operatorname{op}} + Q_{\operatorname{op}} p_{\operatorname{op}} + R_{\operatorname{op}} p_{\operatorname{ac}}) \nonumber\\
\dot{p}_{\operatorname{ac}} &= -2 p_{\operatorname{ac}}( \gamma_{\operatorname{ac}} + Q_{\operatorname{ac}} p_{\operatorname{ac}} + R_{\operatorname{ac}} p_{\operatorname{op}})
\end{align}
Lotka-Volterra systems are well known for modeling the evolution of predator-prey populations. We define the single-mode equilibrium powers $\bar{p}_{\operatorname{op}}$ and $\bar{p}_{\operatorname{ac}}$ as~:
\begin{align}\label{eq:LV-eq}
\bar{p}_{\operatorname{op}} &=  \dfrac{-\gamma_{\operatorname{op}}}{Q_{\operatorname{op}}} &
\bar{p}_{\operatorname{ac}} &=  \dfrac{-\gamma_{\operatorname{ac}}}{Q_{\operatorname{ac}}}
\end{align}
The \emph{effective} linear coefficients are defined by~:
\begin{align*}
	d_{\textrm{op}} &= \gamma_{\textrm{op}} + \bar{p}_{\textrm{ac}} R_{\operatorname{op}} &
	d_{\textrm{ac}} &= \gamma_{\textrm{ac}} + \bar{p}_{\textrm{op}} R_{\operatorname{ac}} 
\end{align*}
And the inter-mode mixing coefficient $\Delta$ is defined by~:
\begin{align*}
	\Delta &= 1 - \dfrac{R_{\operatorname{op}}R_{\operatorname{ac}}}{Q_{\operatorname{op}}Q_{\operatorname{ac}}}
\end{align*}

\begin{figure}[!t]
\centering
\includegraphics[width=\linewidth]{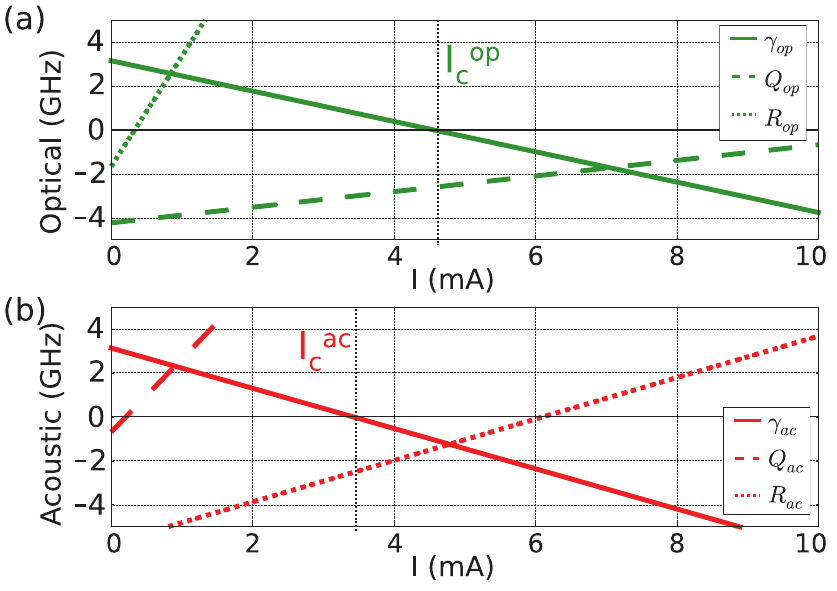}
\caption{Linear and non-linear dissipative coefficients versus applied current $I$ for $H_x=-40$~kA/m and $J_{\operatorname{RKKY}}=-5\times10^{-4}$~J/m$^2$. (a) Optical coefficients, (b) acoustic coefficients. All values are divided by $2\pi$ to be in units of Hz and not in rad/s.}
\label{fig:plot_current}
\end{figure}

The convergence to equilibrium for the LV system is described in reference~\citen{Hofbauer1988} and references~\citen{Bomze1983,Bomze1995} provide a classification with state diagrams. The two-modes system has four equilibriums, their conditions for existence and stability are defined by~:
\begin{itemize}
	\item $\mathcal{P}_0=(0,0)$ if $\gamma_{\operatorname{op}} > 0$ and $\gamma_{\operatorname{ac}} > 0$~:\\
		No mode is excited, this is the subcritical regime with only damped modes.
	\item $\mathcal{P}_\textrm{op}=(\bar{p}_\textrm{op}, 0)$ if $\gamma_{\operatorname{op}} < 0$, $Q_{\operatorname{op}} > 0$ and $d_{\operatorname{ac}} > 0$~:\\
		Only the optical mode is excited and the acoustic mode vanishes.
	\item $\mathcal{P}_\textrm{ac}=(0,\bar{p}_\textrm{ac})$ if $\gamma_{\operatorname{ac}} < 0$, $Q_{\operatorname{ac}} > 0$ and $d_{\operatorname{op}} > 0$~:\\
		Only the acoustic mode is excited and the optical mode vanishes.
	\item $\mathcal{P}^*=(p_\textrm{op}^*, p_\textrm{ac}^*)$ if $d_{\operatorname{op}}Q_{\operatorname{ac}} < 0$, $d_{\operatorname{ac}}Q_{\operatorname{op}} < 0$, $Q_{\operatorname{op}}Q_{\operatorname{ac}}\Delta > 0$ and $(d_{\operatorname{op}}+d_{\operatorname{ac}})/\Delta < 0$~:\\
		The system converges to a mixed-mode equilibrium where both modes have a finite power given by:
\begin{align}
p_{\operatorname{op}}^* &= \dfrac{-d_{\operatorname{op}}}{Q_{\textrm{op}}\Delta} &
p_{\operatorname{ac}}^* &= \dfrac{-d_{\operatorname{ac}}}{Q_{\textrm{ac}}\Delta}
\end{align}
\end{itemize}
Notice that $\mathcal{P}_0$ and $\mathcal{P}^*$ are compatible, they can be stable local equilibriums at the same time, but they are incompatible with $\mathcal{P}_\textrm{op}$ and $\mathcal{P}_\textrm{ac}$. And reciprocally, $\mathcal{P}_\textrm{op}$ and $\mathcal{P}_\textrm{ac}$ can be stable at the same time, but not at the same time as $\mathcal{P}_0$ and $\mathcal{P}^*$.\\
Given the specific conditions are fulfilled, each equilibrium is defined and locally stable. However, the global convergence to this equilibrium depends on the initial conditions, if they are in the basin of convergence of this equilibrium. For instance, $\mathcal{P}_\textrm{op}$ and $\mathcal{P}_\textrm{ac}$ can be stable at the same time, it depends on the initial conditions if the system converges to one or the other equilibrium, or even if it diverges (which corresponds to a switching of one or both layers). See Fig.~4 in reference~\citen{DeAguiar2007} for a phase portrait of $p_\textrm{ac}$ versus $p_\textrm{op}$ \textemdash~noted $n_1$ and $n_2$.\\
\par
The coefficients of Eq.~\eqref{eq:powers} are plotted in figure~\ref{fig:plot_current} versus applied current $I$ for the macrospin parameters defined previously and for $H_x=-40$~kA/m and $J_{\operatorname{RKKY}}=-5\times10^{-4}$~J/m$^2$. Two threshold currents for the modes excitations, $I_c^{\textrm{ac}}$ and $I_c^{\textrm{op}}$, are defined by the vanishing of the linear coefficients $\gamma_{\text{op}}$ and $\gamma_{\text{ac}}$, respectively. For this particular set of parameters, the acoustic threshold current $I_c^{\textrm{ac}}$ is lower than the optical threshold current $I_c^{\textrm{op}}$. Therefore the critical current $I_c$ corresponds to the acoustic threshold current, which is $I_c=3.4$~mA in this particular case. Above the critical current $I_c$, the acoustic mode is excited, and because $Q_{\textrm{ac}}$ is positive (not shown in Figure~\ref{fig:plot_current} above $2$~mA $Q_{\textrm{ac}}$ increases linearly), the power converges to the equilibrium acoustic power; the optical mode remains zero. Above the optical threshold current, the equilibrium acoustic power still exists and it is stable, because $d_{\textrm{op}}>0$ (not shown on the figures). However, $Q_{\textrm{op}}$ is negative, so no equilibrium optical power is defined and the optical mode may diverge. Therefore, the final state depends on the initial conditions~: if the acoustic power is close to the equilibrium $\bar{p}_\textrm{ac}$ and the optical power is close to 0, the system converges to the powers $\{0; \bar{p}_{\textrm{ac}}\}$; if the optical mode diverges faster than the acoustic mode converges to its equilibrium value, the whole system will diverge, which corresponds to a reversal of the layers.\\
\par
Having defined the equilibrium powers, the oscillation frequency is given by the phase equations that will be analyzed in the next section.

\subsection{Phase equations}\label{section:phase-equation}
The corresponding phase equations of Eq.~\ref{eq:coupled_1} including only resonant terms are~:
\begin{align}\label{eq:phase0}
\dot{\phi}_{\operatorname{op}} &= \omega_{\operatorname{op}}+ N_{\operatorname{op}}p_{\operatorname{op}} + T p_{\operatorname{ac}} \nonumber\\
\dot{\phi}_{\operatorname{ac}} &= \omega_{\operatorname{ac}}+ N_{\operatorname{ac}}p_{\operatorname{ac}} + T p_{\operatorname{op}}
\end{align}
We notice that the phase velocities $\dot{\phi}_{\operatorname{op}}$ and $\dot{\phi}_{\operatorname{ac}}$ of the two modes are constant if the powers are at equilibrium ($\dot{p}_{\operatorname{op}}=\dot{p}_{\operatorname{ac}}=0$). Moreover, the phase of each mode depends not only on its own power, but also on the power of the other mode through the non-linear phase mixing $T$. But both phases are independent of each other~: each mode oscillates at its own constant frequency. Note that this is true only if the non-resonant terms are excluded, as shown in section~\ref{section:correction} below.\\

Let's consider the case of a single-mode excitation of the acoustic mode, as it is observed in the simulations shown in this paper. In this case, $p_{\operatorname{op}}=0$ and $p_{\text{ac}}=\bar{p}_{\text{ac}}=\dfrac{-\gamma_{\operatorname{ac}}}{Q_{\operatorname{ac}}}>0$. Therefore the magnetization oscillates at the frequency $f$ of the excited acoustic mode, which is given by~:
\begin{align}\label{eq:frequency}
2\pi f =  \omega_{stt} = \Omega_{\text{ac}} = \omega_{\operatorname{ac}}+ N_{\operatorname{ac}}\bar{p}_{\operatorname{ac}}
\end{align}

\begin{figure}[!t]
\centering
\includegraphics[width=0.9\linewidth]{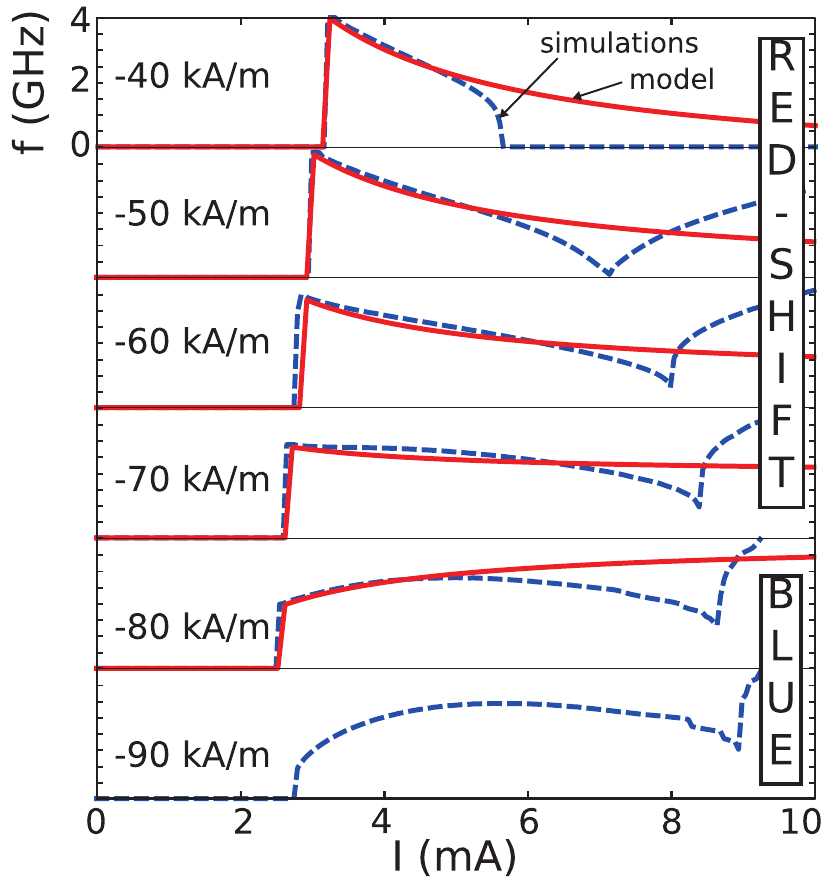}
\caption{Self-sustained oscillations frequency versus applied current $I$ with $J_{\operatorname{RKKY}}=-5\times10^{-4}$~J/m$^2$ and for different applied fields, from top to bottom~: $-40$~kA/m to $-90$~kA/m. The frequency scale is identical in all the panels, from 0 to 4~GHz. Solid red line~: computed from the extended NLAO model. Dashed blue line~: extracted from LLGS simulations. Beyond the spin-flop transition, for $H_x =-90$~kA/m, the extended NLAO model is not applicable.}
\label{fig:plot_freq}
\end{figure}

This equation is equivalent to the phase equation of an STO composed of a single-layer (SL) free layer as described in previous work~\cite{Slavin2009}. The power increases with the applied current, and the frequency decreases or increases depending on the sign of $N_{\operatorname{ac}}$. Figure~\ref{fig:plot_freq} shows the transition between the two regimes, red-shift (frequency decrease with the current) and blue-shift (frequency increase) with $J_{\operatorname{RKKY}}=-5\times10^{-4}$~J/m$^2$, by changing the applied field. The frequency is computed from the extended NLAO model and compared to the frequency obtained from macrospin simulations, both show a transition between red-shift and blue-shift at around $-75$~kA/m. The change of regime with applied field in an asymmetric SyF was already observed numerically~\cite{Gusakova2009} and experimentally~\cite{Houssameddine2010}.\\

\begin{figure}[!t]
\centering
\includegraphics[width=\linewidth]{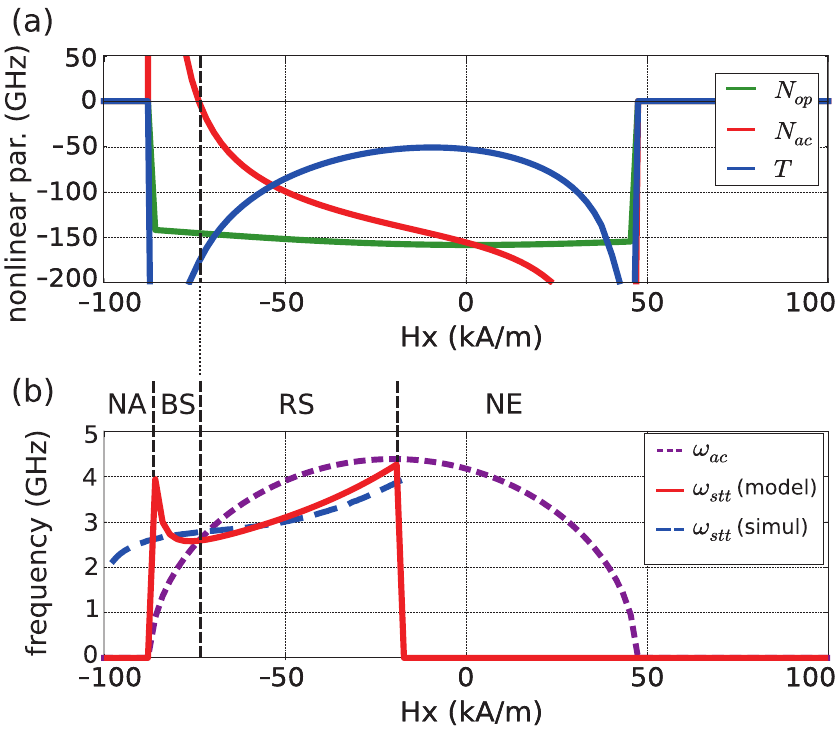}
\caption{(a) Non-linear and (b) linear frequency terms versus applied field $H_x$ for $I=4$~mA and $J_{\operatorname{RKKY}}=-5\times10^{-4}$~J/m$^2$. (a) Non-linear coefficients: optical $N_\text{op}$ (green), acoustic $N_\text{ac}$ (red), inter-mode $T$ (blue). (b) Linear coefficients: dotted magenta line, linear $\omega_\text{ac}$; red solid line, self-sustained oscillations frequency from the model $\omega_{stt}=\Omega_\text{ac}=\omega_\text{ac} + p_{\textrm{ac}} N_{\textrm{ac}}$; dashed blue line, self-sustained oscillations frequency from simulations. The field range is divided in four regions, from low to high fields~: model non-applicable (NA), blue-shift regime (BS), red-shift regime (RS) and no excitation (NE). All values are divided by $2\pi$ to be in Hz units and not in rad/s.}
\label{fig:plot_field}
\end{figure}
As stated, this transition corresponds to $N_{\operatorname{ac}}$ changing sign. The value of the non-linear coefficients of Eq.~\eqref{eq:phase0} is plotted versus applied field $H_x$ in Figure~\ref{fig:plot_field}~(a). $N_{\operatorname{ac}}$ changes signs at around $H_x=-75$~kA/m, which, indeed, corresponds to the red-shift/blue-shift transition. The self-sustained oscillations frequency $f = \omega_{\text{stt}}/(2\pi)$ versus field $H_x$ at $I=4$~mA is reported in Figure~\ref{fig:plot_field}~(b) and compared to the acoustic FMR frequency $\omega_{\text{ac}}$ and the frequency obtained from the simulations. We differentiate four regions, from low to high fields~: (i) below the spin-flop field, at $-90$~kA/m, the extended NLAO model is not valid. (ii) for higher fields but below $-75$~kA/m, the acoustic mode is excited in the blue-shift regime, so $\omega_{\text{stt}} > \omega_{\text{ac}}$. The discrepancy between the frequency obtained from the extended NLAO model and the simulation is high, as expected because the model is not valid anymore if $\omega_{\text{ac}}$ is small. (iii) above $-75$~kA/m, the acoustic mode is excited, in the red-shift regime, $\omega_{\text{stt}} < \omega_{\text{ac}}$. The frequency computed from the model agrees with the simulations. (iv) above $-20$~kA/m, the applied current is too low to excite a mode, the oscillator is in sub-critical mode. Notice that in the vicinity of the field value at which $N_{\operatorname{ac}}$ vanishes, the oscillator frequency does not change much with the applied field, in agreement with the simulations. At this functioning point, the oscillator frequency is not very sensitive neither to the applied field, nor to the applied current.\\
\par
We showed that the frequency of the self-sustained oscillations can be predicted by the extended NLAO model, in the next section the model will be compared to numerical simulations to define its validity range.

\subsection{Single-mode description of the SyF-STO}\label{section:singlemode}

\begin{figure}[!t]
	\centering
	\includegraphics[width=\linewidth]{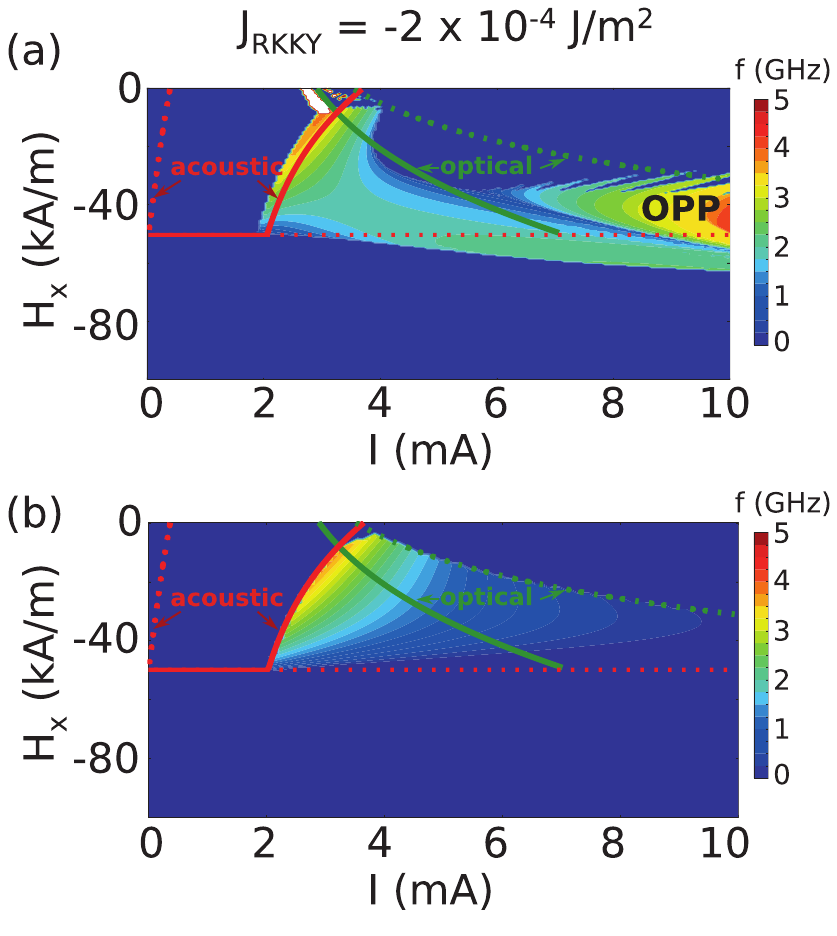}
	\caption{Frequency of the self-sustained oscillations versus applied current and field (a) from macrospin numerical simulations and (b) from the formulas for the power and phase from Eq.~\eqref{eq:frequency}. The RKKY coupling is of $-2\times10^{-4}$~J/m$^2$. Red (green) solid lines represent $I_c(H_x)$ the vanishing of the acoustic (optical) linear dissipative coefficient $\gamma_{\text{ac}(\text{op})}$. Dotted lines correspond to the vanishing of the quadratic dissipative coefficient $Q_{\text{ac}(\text{op})}$.}
	\label{fig:diag_Jex[-2e-4]}
\end{figure}

\begin{figure}[!t]
	\centering
	\includegraphics[width=\linewidth]{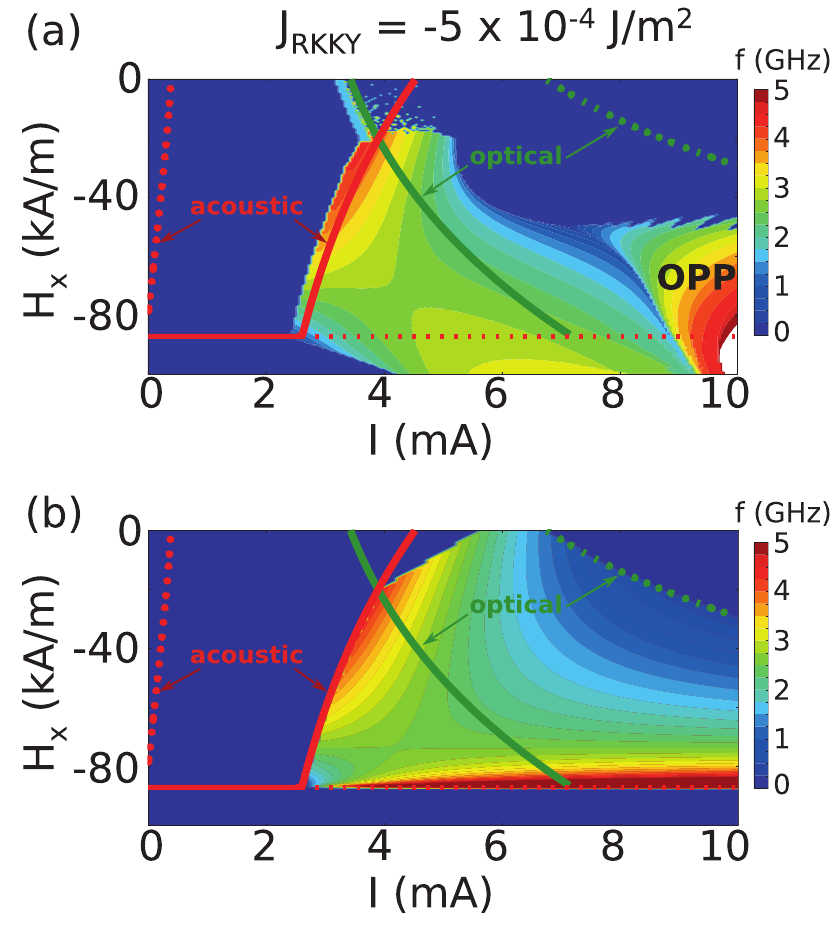}
	\caption{Same as Figure~\ref{fig:diag_Jex[-2e-4]} with an RKKY coupling of $-5\times10^{-4}$~J/m$^2$.}
	\label{fig:diag_Jex[-5e-4]}
\end{figure}

Two sets of simulations are presented, showing the self-sustained oscillations frequency versus applied current and field for two coupling strengths: (i) Figure~\ref{fig:diag_Jex[-2e-4]} in the \emph{small} coupling regime $J_\text{RKKY}=-2\times10^{-4}$~J/m$^2$, (ii) Figure~\ref{fig:diag_Jex[-5e-4]} the \emph{large} coupling regime $J_\text{RKKY}=-5\times10^{-4}$~J/m$^2$. In both figures, the frequency of the $m_{1\,y}$ component of the magnetization of layer 1 from macrospin simulations is plotted in the top panels (a). The frequency computed from the extended NLAO model is plotted in the bottom panels (b). State diagrams for these values of $J_\text{RKKY}$ are displayed in reference~\citen{Monteblanco2017}.\\
We observe a qualitative agreement between the model and the simulations, especially in the region close to the critical current. First, above the acoustic critical current $I_c^{\textrm{ac}}$ (region on the right of the red solid line), the model predicts self-sustained acoustic-like oscillations, just like the simulations (and other publications~\cite{Monteblanco2017}). Just above the optical critical current $I_c^{\textrm{op}}$ (region on the right of the green solid line and on the left of the red solid line), there is no oscillation and the two layers switch, as predicted by the equations of the powers.\\
\par
There are also several discrepancies, that will be discussed in the following.
\par
First, the out-of-plane precession (OPP) region is not predicted by the model. OPP are oscillations around the energy maximum, which are not considered in this model. To describe the OPP, the projection base for the complex $a$-coordinates should be changed to the out-of-plane axes, instead of the equilibrium in-plane axes, and all the coefficients should be computed again.
\par
Second, according to the simulations, self-sustained oscillations are expected when the field is larger than the spin-flop field. However the extended NLAO model is not valid in the spin-flop region. In fact it is not valid in the vicinity of the spin-flop field either, as it was already mentioned. That is why for $J_\text{RKKY}=-2\times10^{-4}$~J/m$^2$, Figure~\ref{fig:diag_Jex[-2e-4]}, the red-shift/blue-shift transition at around $H_x=-45$~kA/m, is not predicted by the extended NLAO model~: it is too close to the spin-flop field value of $-50$~kA/m. On the contrary, for $J_\text{RKKY}=-5\times10^{-4}$~J/m$^2$, Figure~\ref{fig:diag_Jex[-5e-4]}, the red-shift/blue-shift transition at around $H_x=-70$~kA/m, with a spin-flop field at $-90$~kA/m, is well predicted by the extended NLAO model.
\par
Last, the model predicts a much larger region of oscillations than the simulations. In the region on the right of the optical critical current $I_c^{\textrm{op}}$ (green solid line), the difference between the model and the simulations becomes really important. This was also shown in Figure~\ref{fig:plot_freq}. In this region, the power is large, which is a known limit for the validity of the NLAO model. But there could be another explanation, because in this region, the model predicts a single-mode excitation with $p_{\textrm{op}}=0$, whereas the simulations show that $p_{\textrm{op}}$ does not vanish (not shown in the figures).\\
To explain the failure of the model in this region, we propose to study the influence of other terms that we first discarded in the model, namely the linear coefficients from the dissipative part that are non-resonant. The linear terms are important corrections as they depend linearly in the powers, contrary to higher order terms. Also they can be easily computed, which is not the case of higher order terms.

\section{Correction due to non-resonant terms}\label{section:correction}
As was shown in Section~\ref{section:singlemode}, Eq.~\ref{eq:coupled_1} cannot capture all the features of the dynamics, in particular the frequency versus current. Therefore, in order to obtain a better description of the phase, we also include in Eq.~\ref{eq:coupled_1} non-resonant, off-diagonal terms. This leads to the following equation: 
\begin{align}\label{eq:coupled_2}
\dot{b}_{\operatorname{op}} &= -(i\Omega_{\operatorname{op}} + \Gamma_{\operatorname{op}}) b_{\operatorname{op}} - \tilde{\gamma}_{\operatorname{op}}b_{\operatorname{op}}^{\dagger} - \tilde{\vartheta}_{\text{op}}b_{\text{ac}} - \vartheta_{\text{op}}b_{\text{ac}}^{\dagger}\nonumber\\
\dot{b}_{\operatorname{ac}} &= -(i\Omega_{\operatorname{ac}}+ \Gamma_{\operatorname{ac}}) b_{\operatorname{ac}}  -
 \tilde{\gamma}_{\operatorname{ac}} b_{\operatorname{ac}}^{\dagger} - \tilde{\vartheta}_{\text{ac}}b_{\text{op}} - \vartheta_{\text{ac}}b_{\text{op}}^{\dagger}
\end{align}
Here $\Gamma_{\text{op}} = \gamma_{\text{op}} + Q_{\text{op}}p_{\text{op}} + R_{\text{op}}p_{\text{ac}}$ is the optical dissipative part with only resonant terms from equation~\eqref{eq:dissip_final}. Identically, $\Gamma_{\text{ac}} = \gamma_{\text{ac}} + Q_{\text{ac}}p_{\text{ac}} + R_{\text{ac}}p_{\text{op}}$ is the acoustic resonant dissipative part. For the conservative part, $\Omega_{\text{op}} = \omega_{\operatorname{op}}+ N_{\operatorname{op}}p_{\operatorname{op}} + T p_{\operatorname{ac}}$ and $\Omega_{\text{ac}} = \omega_{\operatorname{ac}}+ N_{\operatorname{ac}}p_{\operatorname{ac}} + T p_{\operatorname{op}}$.\\
The coefficients of the non-resonant terms ($\tilde{\gamma}_{\textrm{op}}, \tilde{\gamma}_{\textrm{ac}},\vartheta_{\textrm{op}},\vartheta_{\textrm{ac}},\tilde{\vartheta}_{\textrm{op}},\tilde{\vartheta}_{\textrm{ac}} $) are independent of the powers; for simplicity, we take the coefficients to be real, but taking into account the imaginary part does not change the general conclusions.\\
The equations for the amplitude and phase rewrite as~:
\begin{align}
\dot{p}_{\operatorname{op}} &= -2(\Gamma_{\operatorname{op}} + \tilde{\gamma}_{\operatorname{op}}\cos(2\phi_\text{op})) p_{\operatorname{op}} \nonumber\\
& -  2\sqrt{p_{\text{op}}p_{\text{ac}}}\left(\tilde{\vartheta}_{\text{op}}\cos(\phi_\text{op}-\phi_\text{ac}) + \vartheta_{\text{op}}\cos(\phi_\text{op}+\phi_\text{ac})\right) \nonumber\\
\dot{p}_{\operatorname{ac}} &= -2(\Gamma_{\operatorname{ac}} + \tilde{\gamma}_{\operatorname{ac}}\cos(2\phi_\text{ac})) p_{\operatorname{ac}} \nonumber\\
& -  2\sqrt{p_{\text{op}}p_{\text{ac}}}\left(\tilde{\vartheta}_{\text{ac}}\cos(\phi_\text{op}-\phi_\text{ac}) + \vartheta_{\text{ac}}\cos(\phi_\text{op}+\phi_\text{ac})\right)  \nonumber\\
&\\
\dot{\phi}_{\text{op}} &= \Omega_{\text{op}} + \tilde{\gamma}_{\operatorname{op}}\sin(2\phi_\text{op}) \nonumber\\ &\quad +\sqrt{\dfrac{p_{\operatorname{ac}}}{p_{\operatorname{op}}}} \left(\tilde{\vartheta}_{\text{op}} \sin(\phi_{\text{op}}- \phi_{\text{ac}}) +\vartheta_{\text{op}} \sin(\phi_{\text{op}}+ \phi_{\text{ac}})\right) \nonumber\\
\dot{\phi}_{\text{ac}} &= \Omega_{\text{ac}} + \tilde{\gamma}_{\operatorname{ac}}\sin(2\phi_\text{ac}) \nonumber\\ &\quad +\sqrt{\dfrac{p_{\operatorname{op}}}{p_{\operatorname{ac}}}} \left(\tilde{\vartheta}_{\text{ac}} \sin(\phi_{\text{ac}}- \phi_{\text{op}}) +\vartheta_{\text{ac}} \sin(\phi_{\text{op}}+ \phi_{\text{ac}})\right)
\end{align}

The equations including the non-resonant terms are more complicated, therefore each term will be treated separately.\\
We first present a qualitative interpretation of each term and then evaluate its effect on the dynamics in Figure~\ref{fig:simuls}~: LLGS equation (Eq.~\eqref{eq:LL}), extended NLAO model with only resonant terms (Eq.~\eqref{eq:coupled_1}), with the addition of the linear dissipative terms (Eq.~\eqref{eq:coupled_2}) and with all the terms.

\par
\subsection{Inter-mode phase locking}
An important disagreement between the LLGS simulation (Eq.~\eqref{eq:LL}) and equation~\eqref{eq:coupled_1} is the phase of the non-excited mode, as can be seen from the comparison of Fig.~\ref{fig:simuls}~(a) and~\ref{fig:simuls}~(c). In section~\ref{section:phase-equation}, it was shown that without the non-resonant terms the two modes have different frequencies. However, in the LLGS simulations, Fig.~\ref{fig:simuls}~(a), the two modes are locked, they have the same frequency (although they can have an opposite sign~\cite{Romera2016}). This discrepancy can be corrected by including the terms with the $\tilde{\vartheta}_{\textrm{ac}}$ and $\tilde{\vartheta}_{\textrm{op}}$ coefficients, as is shown in Fig.~\ref{fig:simuls}~(d).\\
\par
Let's suppose that only an acoustic-like mode is excited, but the optical mode does not vanish totally ($p_{\operatorname{op}}\approx0$ and $p_{\operatorname{ac}}=\bar{p}_{\operatorname{ac}}$). The powers are considered to be constant.\\
The differential equation for the phases of the two modes are~:
\begin{align}
	\dot{\phi}_{\text{op}} &= \Omega_{\text{op}} + \sqrt{\dfrac{p_{\operatorname{ac}}}{p_{\operatorname{op}}}} \tilde{\vartheta}_{\text{op}} \sin(\phi_{\text{op}}- \phi_{\text{ac}})\label{eq:dotphi_op}\\
	\dot{\phi}_{\text{ac}} &= \Omega_{\text{ac}} + \sqrt{\dfrac{p_{\operatorname{op}}}{p_{\operatorname{ac}}}} \tilde{\vartheta}_{\text{ac}} \sin(\phi_{\text{ac}}- \phi_{\text{op}})\label{eq:dotphi_ac}
\end{align}
In the acoustic phase equation~\eqref{eq:dotphi_ac}, the second term is negligible compared to the constant frequency $\Omega_{\text{ac}}$ because of the powers ratio, so in the first order, the acoustic mode has a constant frequency $\dot{\phi}_{\text{ac}} = \Omega_{\text{ac}}$, so $\phi_{\text{ac}} = \Omega_{\text{ac}} t$. However, in the optical phase equation~\eqref{eq:dotphi_op}, the second term on the right-hand-side is dominant, also with respect to the left-hand-side. This leads to the relation $\sin(\phi_{\text{op}}- \phi_{\text{ac}})=\sqrt{\dfrac{p_\text{op}}{p_\text{ac}}}\left(\dfrac{\dot{\phi}_{\text{op}} - \Omega_{\text{op}}}{\tilde{\vartheta}_{\text{op}}}\right)\approx 0$, so in the first order,
$\phi_{\text{op}} \approx \Omega_{\text{ac}} t$, or $\phi_{\text{op}} \approx \pi + \Omega_{\text{ac}} t$. This means that the frequency of the non-excited mode is locked to the frequency of the excited mode in the supercritical regime.\\
At the second order, the phase difference is given approximately by~:
\begin{align}
	\phi_{\text{op}} - \phi_{\text{ac}} &= \sqrt{\dfrac{p_\text{op}}{p_\text{ac}}}\left(\dfrac{\Omega_{\text{ac}} - \Omega_{\text{op}}}{\tilde{\vartheta}_{\text{op}}}\right) + k\pi \quad \mbox{with } k\in\mathbb{Z}
\end{align}
\\
\par
Similarly, the terms with the $\vartheta_{\textrm{ac}}$ and $\vartheta_{\textrm{op}}$ coefficients are responsible for a locking with opposite frequency (same absolute frequency, but opposite phase sign), of the form~: $\phi_{\text{op}} + \phi_{\text{ac}}\approx0 $, with $\phi_{\text{ac}}(t) = \Omega_{\text{ac}} t$.\\

If both $\tilde{\vartheta}_{\textrm{op}}$ and $\vartheta_{\textrm{op}}$ are included simultaneously, there is a competition between the two terms for the locking of the non-excited mode, to the same or the opposite frequency as the excited mode. The resulting relation between the two phases is more complicated then. However, regarding the time-average of the frequency, the non-excited mode is locked to the frequency of the excited mode if $\lvert\tilde{\vartheta}_{\textrm{op}}\rvert > \lvert\vartheta_{\textrm{op}}\rvert$, and to the opposite frequency if $\lvert\tilde{\vartheta}_{\textrm{op}}\rvert < \lvert\vartheta_{\textrm{op}}\rvert$. In other words, the coefficient with the highest value (in norm) determines the type of locking, direct or opposite. An example for opposite frequency locking is the self-polarized configuration discussed in reference~\citen{Romera2016}.
\par
\subsection{Power oscillations and second harmonics}
Second, let's focus on the term with the $\tilde{\gamma}_{\text{ac}}$ coefficient (it will be similar for the term in $\tilde{\gamma}_{\text{op}}$). We consider a \emph{pure} single-mode excitation of the acoustic mode, so $p_{\textrm{op}}=0$. Note that this analysis is valid for any single-mode non-linear oscillator equation, including the SL case.\\
Without the other non-resonant terms, the power and phase equations of the acoustic mode write~:
\begin{align*}
\dot{p}_{\operatorname{ac}} &= -2\Gamma_{\operatorname{ac}} p_{\operatorname{ac}} -2 \tilde{\gamma}_{\operatorname{ac}}\cos(2\phi_\text{ac}) p_{\operatorname{ac}} \\
\dot{\phi}_{\text{ac}} &= \Omega_{\text{ac}} + \tilde{\gamma}_{\operatorname{ac}}\sin(2\phi_\text{ac})
\end{align*}
In the assumption that the perturbation due to the $\tilde{\gamma}_{\text{ac}}$ term is small, one can use Lindstedt's series to solve this system of equations~\cite{Lacoste2013}. If $\epsilon = \dfrac{\tilde{\gamma}_{\text{ac}}}{\Omega_{\text{ac}}}$ is small, then the power $p_{\text{ac}}$ and phase $\phi_{\text{ac}}$ can be written as power series of $\epsilon$~: $p_{\text{ac}} = p_0 + \epsilon p_1$ and $\phi_{\text{ac}} = \phi_0 + \epsilon \phi_1$. In the zeroth order, $p_0=\bar{p}_{\operatorname{ac}}$ and $\phi_0 =\bar{\Omega}_{\text{ac}} t$, with $\bar{\Omega}_{\text{ac}} = \omega_\text{ac} + N_\text{ac}\bar{p}_{\operatorname{ac}}$. In the first order, the equation for the power deviation $p_1$ and phase deviation $\phi_1$ are~:
\begin{align*}
\dot{p}_1 &= -2\bar{p}_{\text{ac}} Q_{\text{ac}} p_1 - 2\bar{p}_{\text{ac}}\bar{\Omega}_{\text{ac}} \cos(2 \bar{\Omega}_\text{ac} t) \\
\dot{\phi}_1 &= N_\text{ac} p_1 + \bar{\Omega}_{\text{ac}}\sin(2 \bar{\Omega}_\text{ac} t)
\end{align*}
We use the fact that $\bar{p}_{\text{ac}} Q_{\text{ac}} = - \gamma_{\text{ac}} \ll \bar{\Omega}_\text{ac}$, so the first term on the right-hand side of the power equation is neglected. Therefore, in the first order and in the permanent regime, the power $p_{\text{ac}}$ writes~: 
\begin{align*}
p_{\operatorname{ac}}(t) &= \bar{p}_{\operatorname{ac}}\left(1
-\dfrac{\tilde{\gamma}_{\operatorname{ac}}}{\Omega_\text{ac}}\sin(2\bar{\Omega}_\text{ac} t) \right)
\end{align*}
Up to the first order, the phase $\phi_\text{ac}$ is given by~:
\begin{align*}
\phi_{\operatorname{ac}}(t) &= \bar{\Omega}_{\text{ac}} t - \dfrac{\tilde{\gamma}_{\operatorname{ac}}\omega_\text{ac}}{2\bar{\Omega}_\text{ac}^2}\cos(2\bar{\Omega}_\text{ac} t)
\end{align*}

Therefore the term $\tilde{\gamma}_{\text{ac}}$ gives rise to oscillations of the power but also a second harmonics in the frequency spectrum. As a consequence, it also contributes to the STO synchronization by an AC current on the second harmonics. Notice that this term is also present in STO based on a SL free layer but was omitted in previous descriptions~\cite{Slavin2008}.\\

\subsection{Simulations and trajectories}

\begin{figure*}[t]
\centering
\includegraphics[width=0.9\textwidth]{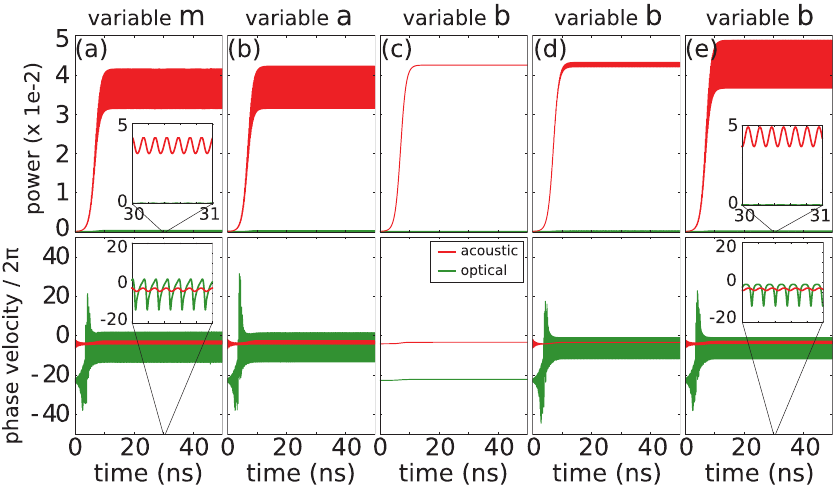}
\caption{Simulations for $H_x=-40$~kA/m, $I=4$~mA and $J_\text{RKKY}=-5\times10^{-4}$~J/m$^2$ performed in the (a) $m$-variables, (b) $a$-variables, (c) $b$-variables, only with resonant terms from Eq.~\eqref{eq:coupled_1}, (d) $b$-variables with non-resonant terms from Eq.~\eqref{eq:coupled_2} but $\tilde{\gamma}_{\operatorname{op}}=\tilde{\gamma}_{\operatorname{ac}}=0$ and (e) $b$-variables with all non-resonant terms from Eq.~\eqref{eq:coupled_2}. The results of the simulations are transformed to the $b$-coordinates to compare them easily. Insets in (a) and (e)~: zoom between 30 and 31~ns. Top panel figures~: powers $p_\text{ac}$ (red) and $p_\text{op}$ (green). Bottom panel figures~: phase velocity or instantaneous frequency in GHz, $\dfrac{\partial\phi_\text{ac}}{\partial t}$ (red) and $\dfrac{\partial\phi_\text{op}}{\partial t}$ (green).}
\label{fig:simuls}
\end{figure*}

\par
The effect of the non-resonant terms on the dynamics is best seen by simulating the different equations.\\

On Fig.~\ref{fig:simuls}, we compare the simulations of different equations and performed in different coordinate systems, and projected afterwards in the $(p_{\text{op}}, p_{\text{ac}}, \phi_{\text{op}}, \phi_{\text{ac}})$-coordinates for comparison. In Fig.~\ref{fig:simuls}~(a), the simulation is performed in the $(m_{1\,x}, m_{1\,y}, m_{1\,z}, m_{2\,x}, m_{2\,y}, m_{2\,z})$-coordinates, like the usual LLGS simulations, according to Eq.~\eqref{eq:LL}.\\
In Fig.~\ref{fig:simuls}~(b), the simulation is performed in the complex $a$-coordinates, using equation~\eqref{eq:form-a-complete}. The trajectory is very similar to the LLGS trajectory. That is because the terms of order superior to 3 in $(a_1, a_2)$ were dropped after the canonical transformation $(\bm{m}_1, \bm{m}_2)\longrightarrow(a_1, a_2)$ and with powers of the order of $10^{-2}$, this approximation is perfectly valid.\\
In Fig.~\ref{fig:simuls}~(c-e), the simulations are performed in the complex $b$-coordinates, from Equation~\eqref{eq:coupled_2}. In Fig.~\ref{fig:simuls}~(c), all the off-diagonal terms are omitted (which corresponds to Eq.~\eqref{eq:coupled_1}). The trajectory exhibits a constant finite acoustic power, a vanishing optical power, and instantaneous frequency for the acoustic and optical mode being constant but with different values. The constant power and frequency of the acoustic mode are close to the averaged values computed from the LLGS equation.\\
In Fig.~\ref{fig:simuls}~(d), $\vartheta_{\text{op}}$, $\vartheta_{\text{ac}}$, $\tilde{\vartheta}_{\text{op}}$ and $\tilde{\vartheta}_{\text{ac}}$ are taken into account. The powers are very similar to the powers obtained in Fig.~\ref{fig:simuls}~(c), which justifies the approximation of constant powers used in the previous section. The frequency of the non-excited mode, the optical mode, is locked to the acoustic frequency. The optical frequency is not constant though, this is because of the competition between the two types of locking, direct and opposite. But its average value is close to the value of the acoustic frequency.\\
In Fig.~\ref{fig:simuls}~(e), $\tilde{\gamma}_{\text{op}}$ and $\tilde{\gamma}_{\text{ac}}$ are also included, so the simulated equation is exactly Eq.~\eqref{eq:coupled_2}. The powers are not constant anymore, but oscillate around the average value instead. Although the average acoustic power is over-estimated compared to the LLGS equation (0.043 instead of 0.036, $20\%$ over-estimated), the average frequencies match more accurately (-3.40~GHz instead of 3.46~GHz, $2\%$ under-estimated).
\par
In conclusion, Eq.~\eqref{eq:coupled_1} with only resonant terms predicts accurately the excitation of the acoustic mode for this set of parameters and it gives a good estimation for the average values of the power and frequency of the excited mode. In order to account for second order features, like phase locking of the non-excited mode to the excited mode and first harmonic oscillation, the corrected equation~\eqref{eq:coupled_2} should be used. However, this corrected model is not enough to explain the discrepancy with the LLGS equation in the average power. When the field becomes closer to the spin-flop field, this error becomes so large that extended NLAO model is not valid anymore. As stated in section~\ref{section:transform}, the error is probably due to higher order terms but this is out of the scope of this paper. Similarly, the extended NLAO model fails at large applied currents and this cannot be explained by the correction terms. It is also probably due to higher order terms.\\
\par
With the restrictions of the model of Eq.~\eqref{eq:coupled_1} in mind, in the next section we make predictions on how to reduce the generation linewidth of the SyF-STO, which is a very important parameter for application. The value of the linewidth given by the model were compared to LLGS simulations.

\section{Application: reduce the STO linewidth}
\subsection{Thermal noise}
So far, the system was supposed to be at zero temperature, however stochastic fluctuations arise at non-zero temperature. The effect of these fluctuations can be estimated in regions where the single-mode approximation is valid. We consider single-mode acoustic-like self-sustained oscillations, but the same reasoning apply to any single-mode non-linear oscillator.\\
With finite temperature, the power and phase of the oscillator are given by~:
\begin{align}
	\dot{p}_{\operatorname{ac}} &= -2 p_{\operatorname{ac}}( \gamma_{\operatorname{ac}} + Q_{\operatorname{ac}} p_{\operatorname{ac}}) + \sqrt{4p_{\textrm{ac}} D_\text{ac}} \;\eta_{p}\\
	\dot{\phi}_{\operatorname{ac}} &= \omega_{\operatorname{ac}}+ N_{\operatorname{ac}}p_{\operatorname{ac}} +\sqrt{\dfrac{D_\text{ac}}{p_{\textrm{ac}}}}\;\eta_{\phi}
\end{align}
Where $\eta_{p}$ and $\eta_{\phi}$ represent white Gaussian noise with normalized variance and the diffusion coefficient $D_\text{ac}$ is defined by~:
\begin{align*}
	D_\text{ac}&=\Gamma_\text{ac}^+ \dfrac{\omega_T}{\Omega_\text{ac}} &\textrm{with}\quad \omega_T&=\dfrac{\gamma_0 k_B T}{2 \mathcal{M}}
\end{align*}
with $\Gamma_\text{ac}^+$ the positive damping (without the contribution from the STT) computed at $\bar{p}_\text{ac}$ and $\Omega_\text{ac} = \omega_\text{ac}+ N_\text{ac}\bar{p}_\text{ac}$.
\par
Because of the thermal noise, the auto-oscillator exhibits a finite generation linewidth $\Delta \omega$, typical of a non-linear single-mode oscillator~\cite{Kim2008,Tiberkevich2008}. The spectral density can be Lorentzian or Gaussian depending on the value of the damping rate of the power fluctuations (or power relaxation rate) $\Gamma_p = \bar{p}_\text{ac}Q_\text{ac}$. The characterization of a non-linear single-mode oscillator in the presence of thermal noise is detailed in Appendix~\ref{A:thermal}.\\
\par
If the correlation time of the power fluctuations ($1/\Gamma_p$) is small compared to the characteristic phase decoherence time (the inverse of the generation linewidth being a good estimation), $\Delta \omega \ll \Gamma_p$, the spectral density is Lorentzian and the full width at half-maximum (FWHM) $\Delta \omega_{L}$ is given by~:
\begin{align}\label{eq:linewidth}
	\Delta \omega_L &= \Delta \omega_0\left(1+\nu_\text{ac}^2\right)
\end{align}
\begin{align}\label{eq:linear_linewidth}
	\text{with } &\nu_\text{ac}=N_\text{ac}/Q_\text{ac} &\text{and}\quad \quad \Delta \omega_0 &= \Gamma_\text{ac}^+ \dfrac{\omega_T}{\bar{p}_\text{ac}\Omega_\text{ac}} 
\end{align}
Where $\Delta \omega_0$ is the \emph{linear} generation linewidth and $\nu_\text{ac}$ is the normalized non-linear frequency shift coefficient.\\

On the other hand, if the correlation time of the power fluctuations is much larger than the decoherence time, $\Delta \omega \gg \Gamma_p$, the spectral density is Gaussian with standard deviation $\Delta \omega_G$ given by~:
\begin{align}
	\Delta \omega_G &= \lvert \nu_\text{ac} \rvert \sqrt{\Delta \omega_0 \Gamma_p}
\end{align}
The FWHM is given by $\sqrt{8\ln 2}\;\Delta\omega_G $.\\
\par

\subsection{Key parameters to the linewidth }
The expressions of Eq.~\ref{eq:linewidth} and~\ref{eq:linear_linewidth} identify three parameters that can be changed to reduce the value of the linewidth to make functional devices~: (i) increase the power relaxation rate $\Gamma_p$, (ii) decrease the linear linewidth $\Delta\omega_0$ and (iii) decrease the normalized non-linear parameter $\nu_{\text{ac}}$.
\par
\begin{itemize}
	\item The power relaxation rate $\Gamma_p = \bar{p}_\text{ac}Q_\text{ac} = \lvert \gamma_\text{ac}\rvert$, is proportional to the difference between the applied current and the critical current $I_c$. An analytical expression of $\gamma_\text{ac}$ is given in reference~\citen{Lacoste2014}. In order to increase $\Gamma_p$ without increasing $I_c$, the absolute value of the slope of $\lvert \gamma_\text{ac}\rvert$ versus $I$ should be increased without increasing $\lvert \gamma_\text{ac}\rvert$ at $I=0$.

	\item The linewidth is proportional to the square of the normalized non-linear parameter $\nu_{\textrm{ac}}$ (if the normalized non-linear parameter is large, which is the case for STOs). Therefore, one way of reducing the linewidth would be to reduce the non-linear parameter $N_\textrm{ac}$ to zero. For the SyF structure discussed here, this is the case at the transition from the redshift to the blueshift regime. At the transition, the linewidth is equal to the linear linewidth value $\Delta\omega_0$. In SL-STO, the vanishing of the non-linear parameter can be achieved by changing the equilibrium magnetic state from in-plane along the easy axis to in-plane along the hard axis or out-of-plane~\cite{Slavin2008}. This usually requires an external field. In SyF-STO, the vanishing of $\nu_{\text{ac}}$ can be achieved by applying an in-plane magnetic field along the easy axis. Such a magnetic field can be generated by the dipolar field from another magnetic layer with the same easy axis direction. Notice that a vanishing non-linear parameter $N_\textrm{ac}$ means that the frequency of the STO becomes independent of its power, and then of the applied current; this loss of tunability can be detrimental for applications. The synchronization bandwidth with an external signal is also proportional to the normalized non-linear parameter $\nu_{\textrm{ac}}$~\cite{Slavin2009}, so it should not be too small.
		
	\item The linear linewidth is inversely proportional to the geometrical mean magnetic volume $\mathcal{M}$ (see Eq.~\eqref{eq:linear_linewidth}). With a SL, the critical current is proportional to the magnetic volume, so it is counter-productive to increase it. For a SyF however, one can think of a thin layer subjected to the spin-transfer torque from the reference layer, coupled to a thick layer not subjected to spin transfer torque. Thus the critical current remains low, whereas the mean magnetic volume is increased.
	
\end{itemize}
In the next sections, we give some ideas about improving these three parameters using a SyF-STO.

\subsection{Dependence of $\Gamma_p$ on the coupling strength}
First, we study the variation of the power relaxation rate $\Gamma_p$ with some parameters of the SyF. However, because $\Gamma_p$ is related to the critical current $I_c$, we need to somehow normalize its value. To start, the super-criticality $\zeta$ is used instead of the current~:
\begin{align*}
	\zeta &= \dfrac{I-I_c}{I_c}
\end{align*}
\par
\begin{figure}[!t]
	\centering
	\includegraphics[width=0.95\linewidth]{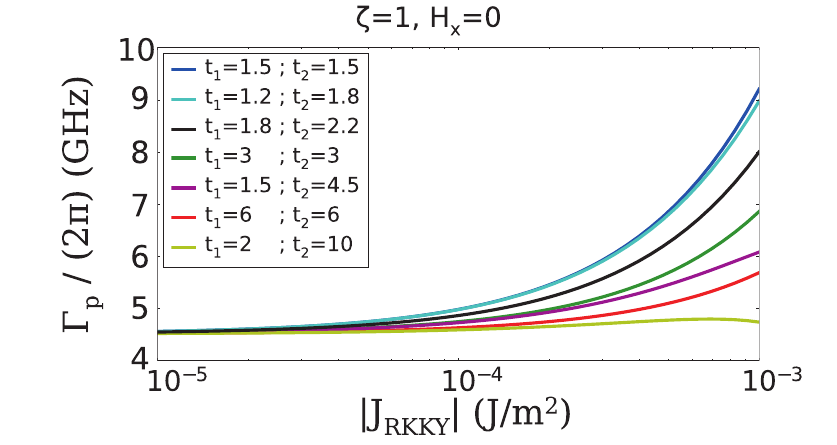}
	\caption{Power relaxation rate $\Gamma_p$ at constant super-criticality $\zeta=1$ and $H_x=0$ versus RKKY coupling energy by area, plotted for different layer thicknesses (in nm). The other layer properties are the same as in Table~\ref{table:macrospin} without applied field, $H_x=0$. $\Gamma_p$ is divided by $2\pi$ to be expressed in Hz instead of rad/s.}
	\label{fig:Gammap}
\end{figure}

Using this normalized quantity, one can compare the values of $\Gamma_p$ at twice the critical current value, which corresponds to $\zeta=1$.\\
The applied field dependence of $\Gamma_p$ is non-trivial but its value at zero field, $H_x=0$, is interesting for applications. The value of $\Gamma_p$ at zero field, for the same super-criticality $\zeta=1$, is plotted in Figure~\ref{fig:Gammap} for different thicknesses of the two layers. It shows that $\Gamma_p$ increases with the RKKY coupling strength, although it remains in the same order of magnitude as with a single layer (asymptotic value for $J_{\textrm{RKKY}}\rightarrow 0$).\\
\par

\subsection{Vanishing of the non-linear parameter~$N_\textrm{ac}$}
\begin{figure}[!t]
\centering
\includegraphics[width=0.95\linewidth]{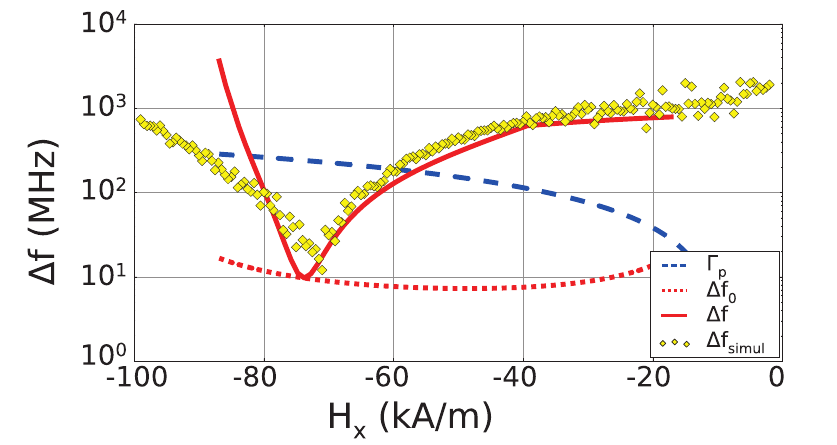}
\caption{Linewidth of $m_{1\;y}$ (yellow diamonds) from LLGS simulations at 300~K, compared to the linewidth (solid red line), \emph{linear} linewidth (dotted red line) and $\Gamma_p$ (dashed blue line) computed from the extended NLAO model, versus applied field for a current of $I = 4$~mA and $J_{\operatorname{RKKY}}=-5\times10^{-4}$~J/m$^2$.}
\label{fig:linewidth}
\end{figure}

\begin{figure}[!t]
\centering
\includegraphics[width=0.95\linewidth]{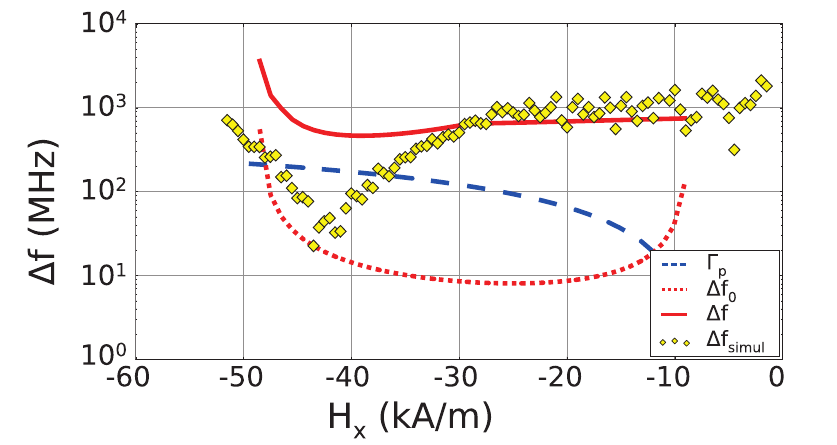}
\caption{Same as Figure~\ref{fig:linewidth} with $J_{\operatorname{RKKY}}=-2\times10^{-4}$~J/m$^2$ and $I = 3$~mA.}
\label{fig:linewidth_lowRKKY}
\end{figure}

Because of the quadratic dependence of the linewidth on the normalized non-linear parameter $\nu_\textrm{ac}$, the most effective action to reduce the oscillator linewidth is to decrease $N_\textrm{ac}$ by applying an in-plane field so the oscillator is excited close to the transition between red-shift and blue-shift.\\
Figure~\ref{fig:linewidth} shows a comparison of the linewidth from LLGS simulations at 300~K and from the extended NLAO model. The linewidth is plotted versus applied field, at $I = 4$~mA and $J_{\operatorname{RKKY}}=-5\times10^{-4}$~J/m$^2$. For the simulations, the linewidth is calculated from a fit to a Lorentzian function. We observe a decrease of the linewidth of almost two orders of magnitude between $H_x=0$ and $H_x=-70$~kA/m. The linewidth decrease is associated to the vanishing of the non-linear parameter $N_{\text{ac}}$. For small fields, $\lvert H_x\rvert < 50$~kA/m, the linewidth is much larger than the power relaxation rate, which corresponds to a Gaussian spectrum. On the other hand, around $H_x=-70$~kA/m, the spectrum has a Lorentzian profile. In the simulations, the spectrum appears to be indeed Lorentzian around $H_x=-70$~kA/m. It is difficult to conclude about the line shape at lower absolute field value, though, because the noise is too large and both profiles interpolate well the simulated spectrum.\\
Figure~\ref{fig:linewidth_lowRKKY} shows the linewidth versus field for a low coupling, $J_{\operatorname{RKKY}}=-2\times10^{-4}$~J/m$^2$, and for $I = 3$~mA. As was shown above, the model does not predict a vanishing of $N_\textrm{ac}$, therefore the predicted linewidth remains large in the whole field range. However, the macrospin simulations show a redshift/blueshift transition at $H_x = -45$~kA/m and a decrease of the linewidth to its \emph{linear} value at this field. In fact, at this field, the frequency does not change with the applied current. In a single-mode model, it means that the phase does not depend on the power, so the linewidth is given by the linear linewidth alone. Therefore, in the low coupling regime, the oscillation looks like it is single-mode, according to the macrospin simulations, but the extended NLAO model is not sufficient to estimate the characteristic parameters of the oscillator.

\subsection{Coupling to a thick layer}

\begin{figure}[!t]
\centering
\includegraphics[width=0.95\linewidth]{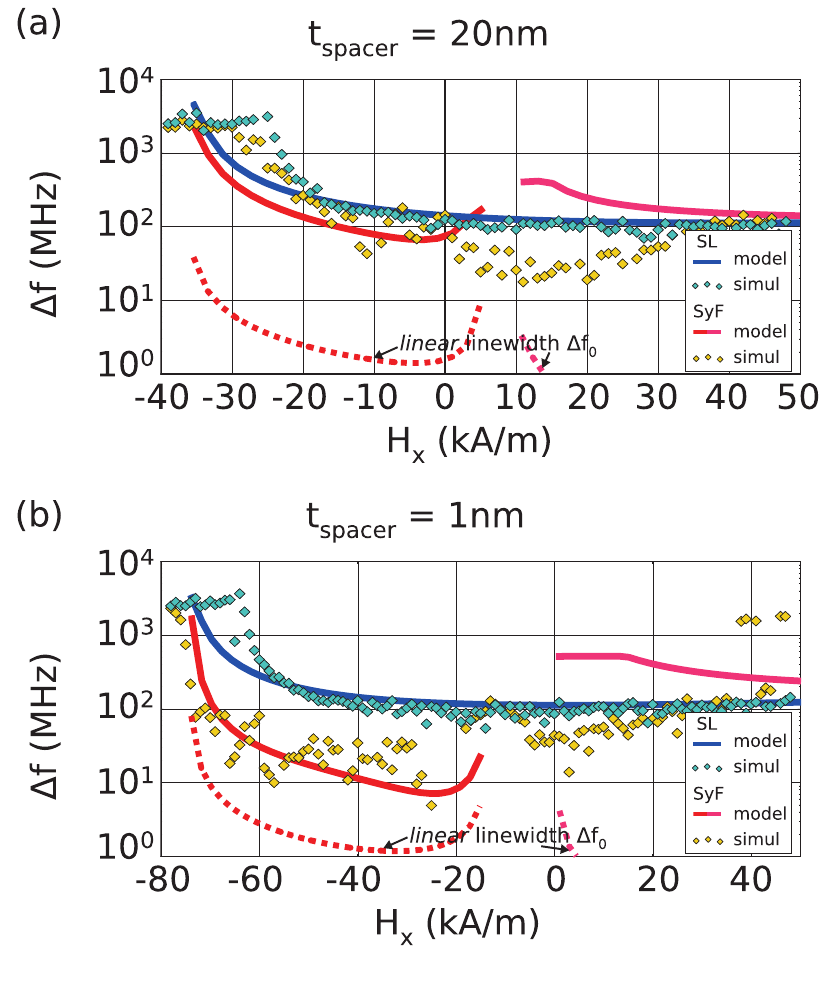}
\caption{Linewidth of $m_{1\,y}$ versus field, comparison between a 2~nm thick single layer (blue) and a 2~nm layer coupled with a 20~nm thick layer (red-orange), separated by (a) 1~nm and (b) 20~nm spacer. Symbols~: LLGS simulations, solid lines~: extended NLAO model, dotted lines~: \emph{linear} linewidth from the extended NLAO model.}
\label{fig:resonator}
\end{figure}
Finally, the last parameter that can be tuned to reduce the linewidth is the linear linewidth $\Delta\omega_0$. The linear linewidth does not depend much on the coupling strength, but more on the magnetic volume, as stated before. In order to increase the total magnetic volume and keep a reasonable critical current, we can imagine a thin layer of 2~nm coupled to a thick layer of 20~nm. With this geometry, where the layers are very asymmetric, the coupling strength plays an important role. Contrary to the asymmetric case studied previously where the non-linear parameter vanishes with the combination of asymmetric layers and strong coupling~\cite{Monteblanco2017}, in the following example, the non-linear parameter is reduced, so the linewidth is decreased, although not to the level of the \emph{linear} linewidth. However, the linewidth reduction happens at lower fields, more suitable for application, than in the previous case.\\
The SyF of this example is compared to a nano-pillar based on a single free layer. The SL-STO is composed of three layer~: (1) a reference layer with in-plane fixed magnetization, a spin polarization of $0.3$ and compensated dipolar fields (the total stray field is zero), (2) a tunnel barrier, and (3) a 2~nm thick free layer, with saturation magnetization of $1\times10^6$~A/m and damping constant of $0.02$. The nano-pillar has an elongated shape of $150\times100$~nm, giving a shape anisotropy to the free layer along the $x$-axis. The SyF-STO of study comprises the same SL nano-pillar, plus two additional layers~: (4) a spacer of variable thickness, 1~nm or 20~nm, and (5) a 20~nm thick free layer with saturation magnetization of $1\times10^6$~A/m and damping constant of $0.02$. The magnetizations of the thick and the thin layers are coupled through dipolar field, whose strength is lower or higher depending on the thickness of the spacer. For the two cases, strong and weak coupling, the coefficients of equation~\eqref{eq:energy} take the values~:\\
\par\text{Strong coupling ($t_{\textrm{MgO}}=1$~nm)~:}
\begin{align*}
(\tilde{D}_x, \tilde{D}_y, \tilde{D}_z)/S &= (-1.9, -2.9, 4.8)\times10^{-4} J/m^2\\
\end{align*}
\par\text{Weak coupling ($t_{\textrm{MgO}}=20$~nm)~:}
\begin{align*}
	(\tilde{D}_x, \tilde{D}_y, \tilde{D}_z)/S &= (-0.9, -1.4, 2.3)\times10^{-4} J/m^2\\
\end{align*}
Due to the shape anisotropy, the magnetization of the thick layer is more stable than that of the thin layer, but it is still free to move. Like for the other stacks studied in this paper, the current is spin polarized between the reference layer and the 2~nm thin layer, but it is considered unpolarized at any other point, including between the thin and thick layers. The simulations were performed at 300~K and the linewidth is computed doing a Lorentzian fit of the power spectral density of $m_{1\,y}$, the magnetization of the thin layer along the $y$-axis. The linewidth computed from the extended NLAO model and extracted from the simulation are showed in Figure~\ref{fig:resonator}.\\
When the thin and thick layers are separated by 20~nm so they are weakly coupled, Figure~\ref{fig:resonator}~(a), the linewidth is of the same order of magnitude with or without the thick layer, in the hundreds of MHz range. Around $10$~kA/m, which is the coupling field (at which the thick and thin layers have the same FMR frequency), the extended NLAO model predicts an increase of the linewidth above the SL value, that is not observed in the simulations. On the contrary, the simulations show a decrease of the linewidth around $10$~kA/m that we cannot explain. Overall, the value of the SyF linewidth is essentially comparable to the value of the SL linewidth.\\
In the strongly coupled case, with a 1~nm spacer, Figure~\ref{fig:resonator}~(b), below the coupling field (around -10~kA/m), the model predicts a reduction of the linewidth of one order of magnitude between the SyF and the SL case; above the coupling field, an increase of the linewidth is predicted. The simulations show a decrease of the linewidth of almost one order of magnitude for the SyF compared to the SL case for fields smaller than the coupling field, with a minimum of 5~MHz at -25~kA/m, in agreement with the model. Notice that the decreased linewidth is still one order of magnitude larger than the \emph{linear} linewidth. Around the coupling field, the SyF and SL linewidths are equivalent, around 100~MHz. Above the coupling field, the linewidth of the SyF is half the linewidth of the SL, in disagreement with the model.\\
In conclusion, we observe a reduction of the linewidth when a thin layer is strongly coupled to a thick layer. The linewidth reduction occurs for all the fields except for the coupling field, at which the linewidth value is as high as for a single layer.

\section{Conclusion}
We presented an extension of the NLAO model to describe the self-sustained oscillations of a SyF composed of two layers coupled with RKKY coupling, dipolar coupling and mutual STT. The analysis was restricted to the plateau region of the SyF, where the two layers are aligned along the same direction at equilibrium, parallel or anti-parallel. However, nothing prevents one from applying the same analysis to arbitrary initial configurations (and an arbitrary number of layers), by taking into account a transverse field for instance, although the diagonalization of the hamiltonian matrix would be more complicated and only numerical solution would be available.
\par
In the extended model, the SyF dynamics is described by two coupled complex non-linear equations, which correspond, in the linear regime, to the acoustic and optical mode. In this paper, we focused on SyFs with fixed external polarizer and, for the set of parameters that we chose, only one mode is excited at a time, the acoustic-like self-oscillation. Therefore, the dynamics can be described by a single mode power and a phase equation, as in the case of a single layer. It means that the self-sustained oscillations are defined by a constant power, resulting from the balance between natural damping and STT. The frequency consists of a linear part and a non-linear part, proportional to the power and to the non-linear frequency shift $N_\textrm{ac}$. Identically, the linewidth of the power spectral density consists of a linear part and a non-linear part.
\par
It was found that with a strong coupling and if the two layers are asymmetric, for instance if they have different thicknesses, the non-linear frequency shift $N_\textrm{ac}$ can be reduced strongly, so the linewidth is also strongly reduced of one order of magnitude. In particular cases, $N_\textrm{ac}$ can even vanish at a given field, which corresponds to a transition between a red-shift and a blue-shift frequency versus current dependency. At this field, the linewidth is reduced to its \emph{linear} linewidth value, which is a reduction of almost two orders of magnitude. The power relaxation rate $\Gamma_p$ was not found to change much compared to the values found for a single layer STO.
\par
This work confirms the robustness of the NLAO model to describe small oscillations of the magnetization around the equilibrium and it shows that it can be extended to several layers. It also presented a relatively simple system to study the interaction between oscillating modes and we hope it can be extended to more general cases.

\acknowledgements
This work was supported by the European Commission under the FP7 program No.~316657 SpinIcur and the FP7 program No.~317950 MOSAIC.

\clearpage
\appendix
\section{Hamiltonian diagonalization~: transformation $a$-$b$}\label{A:transformationT}
The expression of the coefficients of the transformation matrix $T_{ab}$ are given by the 6 angles~: $\phi_j$, $\psi_j$ and $\theta_j$ for $i=(\text{op},\text{ac})$.\\
First the angles $\phi_j$ (for $j = (\text{op},\text{ac})$) are computed~:
\begin{align*}
R_j^{+} = (\mathcal{A}_1+\omega_j)(\mathcal{A}_2+\omega_j) - \mathcal{D}_{12}^2
\end{align*}

\begin{align*}
\operatorname{us}_j &= \mathcal{A}_1 - \omega_j + \mathcal{D}_{12} + \dfrac{\mathcal{C}_{12}\mathcal{D}_{12} - \mathcal{B}_1(\mathcal{A}_2+\omega_j)}
{R_j^{+}}(\mathcal{B}_1+\mathcal{C}_{12}) \\
&\quad + \dfrac{\mathcal{B}_1\mathcal{D}_{12} - \mathcal{C}_{12}(\mathcal{A}_1+\omega_j)}
{R_j^{+}}(\mathcal{B}_2+\mathcal{C}_{12}) \\
\operatorname{uc}_j &= \mathcal{A}_2 - \omega_j + \mathcal{D}_{12} + \dfrac{\mathcal{B}_2\mathcal{D}_{12} - \mathcal{C}_{12}(\mathcal{A}_2+\omega_j)}{R_j^{+}}(\mathcal{B}_1+\mathcal{C}_{12}) \\
&\quad + \dfrac{\mathcal{C}_{12}\mathcal{D}_{12} - \mathcal{B}_2(\mathcal{A}_1+\omega_j)}{R_j^{+}}(\mathcal{B}_2+\mathcal{C}_{12}) \\
\operatorname{un}_j &= \sqrt{\operatorname{us}_j^2 + \operatorname{uc}_j^2} \\
\end{align*}
\begin{align*}
\sin \phi_j &= \dfrac{\operatorname{us}_j}{\operatorname{un}_j}  & \cos \phi_j = \dfrac{\operatorname{uc}_j}{\operatorname{un}_j}\\
\end{align*}

Next the angles $\psi_j$ (for $j = (\text{op},\text{ac})$)~:
\begin{align*}
R_j^{-} = (\mathcal{A}_1-\omega_j)(\mathcal{A}_2-\omega_j) - \mathcal{D}_{12}^2 
\end{align*}
\begin{align*}
\operatorname{vs}_j &= \mathcal{A}_1 + \omega_j + \mathcal{D}_{12} + \dfrac{\mathcal{C}_{12}\mathcal{D}_{12} - \mathcal{B}_1(\mathcal{A}_2-\omega_j)}
{R_j^{-}}(\mathcal{B}_1+\mathcal{C}_{12}) \\
&\quad + \dfrac{\mathcal{B}_1\mathcal{D}_{12} - \mathcal{C}_{12}(\mathcal{A}_1-\omega_j)}
{R_j^{-}}(\mathcal{B}_2+\mathcal{C}_{12}) \\
\operatorname{vc}_j &= \mathcal{A}_2 + \omega_j + \mathcal{D}_{12} + \dfrac{\mathcal{B}_2\mathcal{D}_{12} - \mathcal{C}_{12}(\mathcal{A}_2-\omega_j)}{R_j^{-}}(\mathcal{B}_1+\mathcal{C}_{12}) \\
&\quad + \dfrac{\mathcal{C}_{12}\mathcal{D}_{12} - \mathcal{B}_2(\mathcal{A}_1-\omega_j)}{R_j^{-}}(\mathcal{B}_2+\mathcal{C}_{12}) \\
\operatorname{vn}_j &= \sqrt{\operatorname{vs}_j^2 + \operatorname{vc}_j^2} \\
\end{align*}
\begin{align*}
\sin \psi_j &= \dfrac{\operatorname{vs}_j}{\operatorname{vn}_j}  & \cos \psi_j = \dfrac{\operatorname{vc}_j}{\operatorname{vn}_j}\\
\end{align*}

And finally, the angles $\theta_j$ (for $j = (\text{op},\text{ac})$) are computed~:
\begin{align*}
\mathcal{F}_1 &= \mathcal{A}_1 - \mathcal{B}_1 - \mathcal{C}_{12} \qquad \mathcal{F}_2 = \mathcal{A}_2 - \mathcal{B}_2 - \mathcal{C}_{12}\\
\end{align*}
\begin{align*}
\tanh \theta _j &= -\dfrac{\cos\phi_j( \mathcal{F}_1 - \omega_j) - \sin \phi_j (\mathcal{F}_2 - \omega_j)}
{\cos\psi_j(\mathcal{F}_1 + \omega_j) - \sin \psi_j (\mathcal{F}_2 + \omega_j)} \\
\end{align*}

\section{Coefficients of the dissipative part}\label{A:dissip}
The dissipative part is expressed as a power series in the $a$-coordinates, truncated after the cubic term~:
\begin{align*}
F_{a_i} = \sum_{p,q,r,s} f_{a_i}^{p,q,r,s} {a_1}^p{a_2}^q{a_1^{\dagger}}^r{a_2^{\dagger}}^s \qquad \mbox{for }i=1,2
\end{align*}
We use the following notations~:
\begin{align*}
\nu_1 &= -m \dfrac{\gamma_0}{2\mathcal{M}}\dfrac{\hbar}{2\lvert e\rvert} I \eta_{1} & \nu_2 &= -mn \dfrac{\gamma_0}{2\mathcal{M}}\dfrac{\hbar}{2\lvert e\rvert} I \eta_{2} \\
\nu_{21} &= + \dfrac{\gamma_0}{2\mathcal{M}}\dfrac{\hbar}{2\lvert e\rvert} I \eta_{21} & \nu_{12} &= - \dfrac{\gamma_0}{2\mathcal{M}}\dfrac{\hbar}{2\lvert e\rvert} I \eta_{12}
\end{align*}
\begin{align*}
\kappa_1 &= -m \dfrac{\gamma_0}{2\mathcal{M}}\dfrac{\hbar}{2\lvert e\rvert} I \beta_{1} & 
\kappa_2 &= -mn \dfrac{\gamma_0}{2\mathcal{M}}\dfrac{\hbar}{2\lvert e\rvert} I \beta_{2} \\
\kappa_{21} &= + \dfrac{\gamma_0}{2\mathcal{M}}\dfrac{\hbar}{2\lvert e\rvert} I \beta_{21} &
\kappa_{12} &= - \dfrac{\gamma_0}{2\mathcal{M}}\dfrac{\hbar}{2\lvert e\rvert} I \beta_{12}
\end{align*}
Hence the non-vanishing coefficients of $F_{a_1}$ and $F_{a_2}$ with indices $(p,q,r,s)$ are given by ($i$ is the imaginary unit, $i^2=-1$)~:
\begin{align*}
F_{a_1}&: \\
&(1, 0, 0, 0): \alpha_1\mathcal{A}_1 + 2n\beta\nu_{21} + 2\beta\nu_1 - 2in\beta\kappa_{21} - 2i\beta\kappa_1\\
&(0, 1, 0, 0): \alpha_1\mathcal{D}_{12} - (1+n)\nu_{21} + i(1+n)\kappa_{21}  \\
&(0, 0, 1, 0): \alpha_1\mathcal{B}_1 \\
&(0, 0, 0, 1): \alpha_1\mathcal{C}_{12}  + (1-n)\nu_{21} - i(1-n)\kappa_{21}\\
& \\
&(2, 0, 1, 0): -\alpha_1\beta\mathcal{A}_1 + 2\alpha_1\mathcal{U}_1 - 2n\beta^2\nu_{21} - 2\beta^2\nu_1 \\
&(1, 1, 0, 1): \alpha_1\mathcal{W}_{12}  - 4n\nu_{21} + 4in\kappa_{21} \\
&(0, 2, 0, 1): \alpha_1\mathcal{Z}_{21} + \dfrac{1+n}{2\beta}\nu_{21}  - i\dfrac{1+n}{2\beta}\kappa_{21} \\
&(0, 1, 0, 2): \alpha_1\mathcal{Y}_{21} - \dfrac{1-n}{2\beta}\nu_{21} + i\dfrac{1-n}{2\beta}\kappa_{21} \\
&(1, 1, 1, 0): 2 \alpha_1\mathcal{Z}_{12} + (1+n)\beta\nu_{21} - i\beta(1+n)\kappa_{21} \\
&(1, 0, 1, 1): 2 \alpha_1\mathcal{Y}_{12} - (1-n)\beta\nu_{21} + i\beta(1-n)\kappa_{21} \\
&(2, 0, 0, 1): 3 \alpha_1\mathcal{Z}_{12} + 3\dfrac{1+n}{2}\beta\nu_{21} - i\dfrac{1+n}{2}\beta\kappa_{21} \\
&(2, 1, 0, 0): 3 \alpha_1\mathcal{Y}_{12} - 3\dfrac{1-n}{2}\beta\nu_{21} + i\dfrac{1-n}{2}\beta\kappa_{21} \\
&(1, 0, 2, 0): 3\alpha_1\mathcal{V}_1 \\
&(3, 0, 0, 0): 3\alpha_1\mathcal{V}_1
\end{align*}

\begin{align*}
F_{a_2}&: \\
&(1, 0, 0, 0): \alpha_2\mathcal{D}_{12}  - (1+n)\nu_{12} + i(1+n)\kappa_{12} \\
&(0, 1, 0, 0): \alpha_2\mathcal{A}_{2}  + \dfrac{2n}{\beta}\nu_{12} + \dfrac{2}{\beta}\nu_2  - i\dfrac{2n}{\beta}\kappa_{12} - i\dfrac{2}{\beta}\kappa_2 \\
&(0, 0, 1, 0): \alpha_2\mathcal{C}_{12}  + (1-n)\nu_{12} - i(1-n)\kappa_{12}\\
&(0, 0, 0, 1): \alpha_2\mathcal{B}_2 \\
& \\
&(0, 2, 0, 1): -\dfrac{\alpha_2}{\beta}\mathcal{A}_2 + 2\alpha_2\mathcal{U}_2 - \dfrac{2n}{\beta^2}\nu_{12} - \dfrac{2}{\beta^2}\nu_2 \\
&(1, 1, 1, 0): \alpha_2\mathcal{W}_{12} - 4n\nu_{12} + 4in\kappa_{12} \\
&(2, 0, 1, 0): \alpha_2\mathcal{Z}_{12} + \dfrac{1+n}{2}\beta\nu_{12} - i\dfrac{1+n}{2}\beta\kappa_{12} \\
&(1, 0, 2, 0): \alpha_2\mathcal{Y}_{12} - \dfrac{1-n}{2}\beta\nu_{12} + i\dfrac{1-n}{2}\beta\kappa_{12} \\
&(1, 1, 0, 1): 2\alpha_2\mathcal{Z}_{21} + \dfrac{1+n}{\beta}\nu_{12} - i\dfrac{1+n}{\beta}\kappa_{12} \\
&(0, 1, 1, 1): 2\alpha_2\mathcal{Y}_{21} - \dfrac{1-n}{\beta}\nu_{12} + i\dfrac{1-n}{\beta}\kappa_{12} \\
&(0, 2, 1, 0): 3\alpha_2\mathcal{Z}_{21} + 3\dfrac{1+n}{2\beta}\nu_{12} - i\dfrac{1+n}{2\beta}\kappa_{12} \\
&(1, 2, 0, 0): 3\alpha_2\mathcal{Y}_{21} - 3\dfrac{1-n}{2\beta}\nu_{12} + i\dfrac{1-n}{2\beta}\kappa_{12} \\
&(0, 3, 0, 0): 3\alpha_2\mathcal{V}_2 \\
&(0, 1, 0, 2): 3\alpha_2\mathcal{V}_2
\end{align*}

\section{Thermal noise and Fokker-Planck equation}\label{A:thermal}
Thermal noise is introduced in Eq.~\eqref{eq:coupled_1} in the form~:
\begin{align}\label{eq:A-noise}
\dot{b}_{\text{op}}  + b_{\text{op}}\left(i\Omega_{\text{op}} + \Gamma_{\text{op}}\right) &= \sqrt{2D_\text{op}} \eta_\text{op} \nonumber\\
\dot{b}_{\text{ac}}  + b_{\text{ac}}\left(i\Omega_{\text{ac}} + \Gamma_{\text{ac}}\right) &= \sqrt{2D_\text{ac}} \eta_\text{ac}
\end{align}
The noise amplitudes $D_\text{op}$ and $D_\text{ac}$, also called diffusion coefficients, are not constant and depend on the mode powers~: $D_\text{op}(b_{\text{op}},b_{\text{ac}})$ and $D_\text{ac}(b_{\text{op}},b_{\text{ac}})$, but this dependence is omitted for clarity. They will be determined later. $\Omega_{\text{op}}$, $\Omega_{\text{ac}}$, $\Gamma_{\text{op}}$ and $\Gamma_{\text{ac}}$ are the conservative (for optical and acoustic modes) and the dissipative deterministic coefficients. They also depend on the mode powers. $\eta_\text{op}$ and $\eta_\text{ac}$ are two independent white noise sources with zero mean and correlators given by~:
\begin{align*}
\langle \eta_i(t) \rangle &= 0 \;\text{,} && \text{ for $i$ $\in$ (op, ac)}\\
\langle \eta_i(t)\eta_j(t') \rangle &= 0 \;\text{,} && \text{ for $i,j$ $\in$ (op, ac)$^2$}\\
\langle \eta_i(t)\bar{\eta}_j(t') \rangle &= \delta_{ij}\delta(t-t') \;\text{,} && \text{ for $i,j$ $\in$ (op, ac)$^2$}
\end{align*}

The expressions of the diffusion coefficients are determined by insuring that the equilibrium probability density function (PDF) for the powers and phase reduces to the Boltzmann distribution without applied current~\cite{Slavin2009}. Considering the Stratonovich stochastic differential equation (SDE)~\eqref{eq:A-noise}, the time evolution of the PDF $\mathcal{P}(p_\text{op}, p_\text{ac}, \phi_\text{op}, \phi_\text{ac}, t)$ is given by the following Fokker-Planck (FP) equation~:
\begin{align*}
\dfrac{\partial\mathcal{P}}{\partial t} &- \dfrac{\partial}{\partial p_\text{op}}\left(2p_\text{op}\Gamma_\text{op}\mathcal{P}\right)  -\dfrac{\partial}{\partial p_\text{ac}}\left(2p_\text{ac}\Gamma_\text{ac}\mathcal{P}\right)  \\
&\quad + \dfrac{\partial}{\partial \phi_\text{op}}\left(\Omega_\text{op}\mathcal{P}\right)
+ \dfrac{\partial}{\partial \phi_\text{ac}}\left(\Omega_\text{ac}\mathcal{P}\right) \\
&= \dfrac{\partial}{\partial p_\text{op}}\left(2p_\text{op}D_\text{op}\dfrac{\partial\mathcal{P}}{\partial p_\text{op}}\right) + \dfrac{\partial}{\partial p_\text{op}}\left(\mathcal{P}\dfrac{\partial}{\partial p_\text{op}}(p_\text{op}D_\text{op})\right) \\
&\quad + \dfrac{\partial}{\partial p_\text{ac}}\left(2p_\text{ac}D_\text{ac}\dfrac{\partial\mathcal{P}}{\partial p_\text{ac}}\right) + \dfrac{\partial}{\partial p_\text{ac}}\left(\mathcal{P}\dfrac{\partial}{\partial p_\text{ac}}(p_\text{ac}D_\text{ac})\right) \\
&\quad + \dfrac{D_\text{op}}{2p_\text{op}} \dfrac{\partial^2\mathcal{P}}{\partial \phi_\text{op}^2} + \dfrac{D_\text{ac}}{2p_\text{ac}} \dfrac{\partial^2\mathcal{P}}{\partial \phi_\text{ac}^2}
\end{align*}
Here, we considered that the diffusion coefficients depend only on the mode powers. The terms in the left-hand-side come from the deterministic equation, or drift, whereas the terms in the right-hand-side represent the thermal diffusion. At equilibrium $\left(\dfrac{\partial\mathcal{P}}{\partial t}=0\right)$, the PDF $\mathcal{P}_0$ is a uniform distribution for the phases, so we can remove the last two drift terms of the left-hand-side. Moreover, the second and fourth diffusion terms of the right-hand side should be compensated by two terms of drift that are usually neglected. They arise from the renormalization of the multiplicative noise terms~\footnote{The diffusion coefficients can be renormalized if the SDE is expressed in the It\={o} form, which differs from the Stratonovich form by an extra drift term. The starting point is a stochastic LLGS equation in the Stratonovich form, which is used to describe physical noise. In order to simplify the expression of the noise diffusion terms, one must convert the SDE to the It\={o} form, then simplify the diffusion coefficients, and convert the SDE back to the Stratonovich form. This process adds two extra drift terms that do not balance each other.} (see reference~\cite{Taniguchi2014} where these extra drift terms are included for a SL free layer). The extra drift terms can be incorporated in~\eqref{eq:A-noise} to give the correct equation in Stratonovich form~:
\begin{align}\label{eq:A-noise_2}
\dot{b}_{\text{op}}  + b_{\text{op}}\left(i\Omega_{\text{op}} + \Gamma_{\text{op}}\right) + f_\text{op} b_\text{op} &= \sqrt{2D_\text{op}} \eta_\text{op} \nonumber\\
\dot{b}_{\text{ac}}  + b_{\text{ac}}\left(i\Omega_{\text{ac}} + \Gamma_{\text{ac}}\right) + f_\text{ac} b_\text{ac}  &= \sqrt{2D_\text{ac}} \eta_\text{ac} \nonumber\\
\end{align}
With~:
\begin{align*}
f_\text{op} &= -\dfrac{1}{2p_\text{op}}\dfrac{\partial(p_\text{op} D_\text{op})}{\partial p_\text{op}} \\
f_\text{ac} &= -\dfrac{1}{2p_\text{ac}}\dfrac{\partial(p_\text{ac} D_\text{ac})}{\partial p_\text{ac}}
\end{align*}
Interestingly, these extra drift terms contribute only to the power equations. In particular, they are responsible for the non-zero average power below threshold (when solving $\dot{p}=0$, $p=0$ is not a solution anymore).\\
\par
After eliminating the extra drift terms, the FP equation at equilibrium reduces to~:
\begin{align*}
0 &=  \dfrac{\partial}{\partial p_\text{op}}\left(2p_\text{op}\Gamma_\text{op}^+ \mathcal{P}_0 + 2p_\text{op}D_\text{op}\dfrac{\partial\mathcal{P}_0}{\partial p_\text{op}}\right) \\
& 
+ \dfrac{\partial}{\partial p_\text{ac}}\left(2p_\text{ac}\Gamma_\text{ac}^+ \mathcal{P}_0 + 2p_\text{ac}D_\text{ac}\dfrac{\partial\mathcal{P}_0}{\partial p_\text{ac}}\right)
\end{align*}
Where $\Gamma_\text{op}^+$ and $\Gamma_\text{ac}^+$ are the dissipative terms at zero applied current, i.e. the natural damping.\\
A solution $\mathcal{P}_0(p_\textrm{op}, p_\textrm{ac})$ of the former equation is~:
\begin{align*}
\mathcal{P}_0 &= Z^{-1}\exp\left(-\int_0^{p_\textrm{op}}\dfrac{\Gamma_\text{op}^+}{D_\text{op}}\text{d}p_\text{op} -\int_0^{p_\textrm{ac}}\dfrac{\Gamma_\text{ac}^+}{D_\text{ac}}\text{d}p_\text{ac}\right)
\end{align*}
Where $Z$ is a normalization constant. The equilibrium PDF should correspond to the Boltzmann distribution, which is equal to $Z'^{-1}\exp\left(-\dfrac{E}{k_B T}\right)$, where $Z'$ is another normalization constant, $E$ is the energy of the system as defined in Eq.~\eqref{eq:energy} and $T$ is the temperature. Then the diffusion coefficients are given by~:
\begin{align}
D_\text{op} &= \Gamma_\text{op}^+\; k_B T \left(\dfrac{\partial E}{\partial p_\text{op}}\right)^{-1} = \Gamma_\text{op}^+ \dfrac{\omega_T}{\Omega_\text{op}} \\
D_\text{ac} &= \Gamma_\text{ac}^+\; k_B T \left(\dfrac{\partial E}{\partial p_\text{ac}}\right)^{-1} = \Gamma_\text{ac}^+ \dfrac{\omega_T}{\Omega_\text{ac}}
\end{align}
\par
We now consider the self-oscillation regime with a single-mode excitation of the acoustic mode, with thermal noise. The stochastic differential equation of the power and phase is expressed in the It\={o} form, which is preferred when solving analytically stochastic equations because the solutions are martingales. For clarity, the $ac$ index is dropped on the power $p$ and phase $\phi$~:
\begin{align}
\dot{p} &= -2 p\Big( \gamma_{\operatorname{ac}} + Q_{\operatorname{ac}} p + \tilde{f}_\text{ac}(p)\Big) + \sqrt{4p D_\text{ac}} \eta_{p}  \label{eq:A-power}\\
\dot{\phi} &= \omega_{\operatorname{ac}}+ N_{\operatorname{ac}}p + \sqrt{\dfrac{D_\text{ac}}{p}}\eta_{\phi}
\end{align}
Where $\eta_{p} = \mathcal{R}e(\sqrt{2}\eta_\text{ac}e^{i\phi_\text{ac}})$ and $\eta_{\phi} = \mathcal{I}m(\sqrt{2}\eta_\text{ac}e^{i\phi_\text{ac}})$ are real stochastic variables with zero average and $\langle \eta_{i}(t)\eta_{j}(t')\rangle = \delta_{ij}\delta(t-t')$, for $i,j \in (p, \phi)$. $\tilde{f}_\text{ac} = 2 f_\text{ac}$ is the extra drift term in the It\={o} form, computed from its Stratonovich form and the diffusion coefficients.\\
Due to the extra $\tilde{f}_\textrm{ac}$ term, the stationary power is different from the power $p_0$ without temperature. However, above the threshold, we suppose that the stationary power $\tilde{p}_0$ is close to the zero-temperature value:
\begin{align*}
\tilde{p}_0 &= p_0(1+\delta p_0) &\mbox{with }\delta p_0 \ll 1
\end{align*}
It can be shown that $\delta p_0$ is given by:
\begin{align*}
	\delta p_0 &= -\dfrac{\tilde{f}_\textrm{ac}(p_0)}{p_0 Q_\text{ac}} = \dfrac{\Delta\omega_0}{\Gamma_p} - \dfrac{\nu\Delta\omega_0}{\Omega_0}\dfrac{\Gamma^{+}(p_\infty)}{\Gamma^{+}(p_0)}
\end{align*}
Where $\Delta\omega_0=\dfrac{D_\text{ac}(p_0)}{p_0}$ is the \emph{linear} generation linewidth, $\Gamma_p = p_0 Q_\textrm{ac}$ is the power relaxation rate, $\Omega_0=\omega_\text{ac} + p_0N_\text{ac}$ is the stationary frequency, $\nu = N_\text{ac}/Q_\text{ac}$ is the normalized non-linear frequency shift coefficient and $p_\infty = -\dfrac{\omega_\text{ac}}{N_\text{ac}}$, with positive or negative value. If $N_{\text{ac}}$ is negative, $p_\infty$ corresponds to the maximum oscillation power, for which $\omega_\text{ac} + N_\text{ac} p_\infty=0$.\\
As long as $\Delta\omega_0 \ll \Gamma_p$ and the oscillation frequency $\Omega_0$ is high enough ($\Omega_0\gg \nu\Delta\omega_0$), the effect of the extra drift term can be neglected and $\tilde{p}_0\approx p_0$. \\
\par
Then, we consider fluctuations of the power around the equilibrium power $p_0$ and of the phase around $\phi_0(t) = \Omega_0t$~: $\delta p = p-p_0$ (with $\delta p \ll p_0$) and $\delta\phi=\phi-\phi_0$~:
\begin{align}
\dot{\delta p} &= -2 p_0 \tilde{Q} \delta p + \sqrt{4p_0 D_\text{ac}} \eta_{p}\\
\dot{\delta \phi} &= N_{\operatorname{ac}}\delta p + \sqrt{\dfrac{D_\text{ac}}{p_0}}\eta_{\phi}
\end{align}
Where the effective non-linear relaxation rate coefficient is $\tilde{Q}=Q_\text{ac} + \dfrac{\partial \tilde{f}_\text{ac}}{\partial p_\text{ac}}\bigg|_{p=p_0}$.\\
The correction due to the temperature-dependent term on the non-linear relaxation rate writes as~:
\begin{align*}
	\dfrac{\tilde{Q}}{Q_\text{ac}} -1 &=  \dfrac{\Delta\omega_0}{\Gamma_p} + \nu\Delta\omega_0 \left(\dfrac{2\omega_\text{ac} - \Omega_0}{\Omega_0^2}\right)\dfrac{\Gamma^{+}(p_\infty)}{\Gamma^{+}(p_0)}
\end{align*} 
The same conditions that assured that $\delta p_0\ll 1$ lead to $\tilde{Q}\approx Q_\text{ac}$.
\par
Because the stochastic equations are linear, the power and phase fluctuations are Gaussian processes with zero mean. There are contributions to the linewidth from the phase noise ($\eta_{\phi}$) and from the amplitude noise ($N_{\operatorname{ac}}\delta p$). Note that the other mode, the optical mode, is considered to be subcritical, so its power is almost zero, and in any case much smaller than the power of the acoustic mode. Therefore its contribution to the power spectral density is neglected.\\
The power is a weakly stationary process but the phase is a non-stationary Gaussian random walk. We obtain the expression of the power variance $\Delta p^2 = \langle \delta p^2\rangle$ and the phase variance $\Delta\phi^2 = \langle \delta\phi^2\rangle$~\cite{Tiberkevich2008}~:
\begin{align*}
\Delta p^2 &= p_0^2 \dfrac{\Delta \omega_0}{\Gamma_p}  \\
\Delta\phi^2 &= \Delta\omega_0 \left[ (1+\nu^2)\lvert t\rvert - \nu^2\dfrac{1- e^{-2\Gamma_p\lvert t\rvert }}{2\Gamma_p}\right]
\end{align*}


\end{document}